\documentclass[aps, prl, reprint, amsmath, amssymb, superscriptaddress, footinbib, floatfix]{revtex4-2}

\usepackage{graphicx}
\usepackage{amssymb}
\usepackage{amsmath}
\usepackage{dsfont}
\usepackage{setspace}
\usepackage{newfloat}
\usepackage{mathtools}
\usepackage{siunitx}
\usepackage{bm}
\usepackage{float}
\usepackage{color}
\definecolor{apsHlColor}{RGB}{46,48,146}
\usepackage[colorlinks=true, citecolor=apsHlColor, linkcolor=apsHlColor, urlcolor=apsHlColor]{hyperref}

\begin{document}

\title{Measuring High-Order Phonon Correlations in an Optomechanical
Resonator}

\author{Y. S. S. Patil}
\email{yogesh.patil@yale.edu}
\affiliation{Department of Physics, Yale University, New Haven, Connecticut 06520, USA}
\author{J. Yu}
\affiliation{Department of Physics, Yale University, New Haven, Connecticut 06520, USA}
\author{S. Frazier}
\affiliation{Department of Physics, Yale University, New Haven, Connecticut 06520, USA}
\author{Y. Wang}
\affiliation{Department of Applied Physics, Yale University, New Haven, Connecticut 06520, USA}
\author{K. Johnson}
\affiliation{Department of Physics, Yale University, New Haven, Connecticut 06520, USA}
\author{J. Fox}
\affiliation{Department of Physics, Yale University, New Haven, Connecticut 06520, USA}
\author{J. Reichel}
\affiliation{Laboratoire Kastler Brossel, ENS-Universit\'{e} PSL, CNRS, Sorbonne Universit\'{e}, Coll\`{e}ge de France 24 rue Lhomond, 75005 Paris, France}
\author{J. G. E. Harris}
\email{jack.harris@yale.edu}
\affiliation{Department of Physics, Yale University, New Haven, Connecticut 06520, USA}
\affiliation{Department of Applied Physics, Yale University, New Haven, Connecticut 06520, USA}
\affiliation{Yale Quantum Institute, Yale University, New Haven, Connecticut 06520, USA}

\begin{abstract}
We use single photon detectors to probe the motional state of a superfluid $^4$He resonator of mass $\sim1$ ng.
The arrival times of Stokes and anti-Stokes photons (scattered by the resonator's acoustic mode) are used to measure the resonator's phonon coherences up to the fourth order. 
By post-selecting on photon detection events, we also measure coherences in the resonator when $\leq3$ phonons have been added or subtracted. 
These measurements are found to be consistent with predictions that assume the acoustic mode to be in thermal equilibrium with a bath through a Markovian coupling.
\end{abstract}

\maketitle

\setlength\abovedisplayskip{3pt}
\setlength\belowdisplayskip{3pt}

Cavity optomechanical systems offer a platform for merging the advantageous features of the optical and the acoustic domains. In the last two decades, coherently coupled optical and acoustic resonators have been used to realize a range of quantum technologies including transducers, sensors, repeaters and memories. Quantum optomechanical devices can also be used in gravitational wave detection, tests of quantum mechanics at macroscopic scales, and searches for physics beyond the standard model \cite{meystre2013, aspelmeyer2014, kurizki2015, bowen_milburn_2015,metcalfe2014, carney2020, romeroisart2011, cronin2009, arndt2012}.

To date, most quantum optomechanical devices have operated in a regime where linear equations of motion accurately describe the optical and mechanical modes, the coupling between them, the drives applied to them and the quantum backaction of their readout.
A number of important results have been achieved in this linear regime, including the preparation of mechanical resonators in the ground state, squeezed states, and bipartite entangled states \cite{chan2011, teufel2011, wollman2015, ockeloenkorppi2018}.
However, the mechanical states accessible in this regime offer no quantum advantage for metrology or information processing; nor do these states demonstrate the most striking features of quantum mechanics, such as violations of Bell-type inequalities or nonclassical quasiprobability functions \cite{matthews2016, tan2019, arvidssonshukur2020}.
As a result, it is of considerable interest to develop optomechanical systems that exhibit nonlinearity at the single quantum level.

One approach to attaining single-quantum nonlinearity is to use the measurement backaction of a single photon detector (SPD) \cite{knill2001, scheel2003}.
To date, this approach has been used in the domains of quantum optics, cavity-QED, and optomechanics  \cite{mandel1999, duan2001, kuzmich2003, hacker2019, cohen2015}.
In optomechanical systems, the detection of a single acoustically scattered photon heralds the creation (or annihilation) of a single phonon in the mechanical oscillator.
Such heralded protocols have been used to measure non-classical effects in mechanical resonators with mass $\sim 1$ pg \cite{riedinger2016, hong2017, riedinger2018, marinkovic2018}. 
In devices with mass $\sim 1$ ng, this approach has been used to measure simpler quantum effects (such as sideband asymmetry), and to verify the thermal character of the two-phonon correlations in the resonator \cite{ galinskiy2020}.

\setlength{\abovecaptionskip}{0pt plus 1pt minus 1pt} 
\begin{figure*}
\centering
\includegraphics[width=0.98\textwidth]{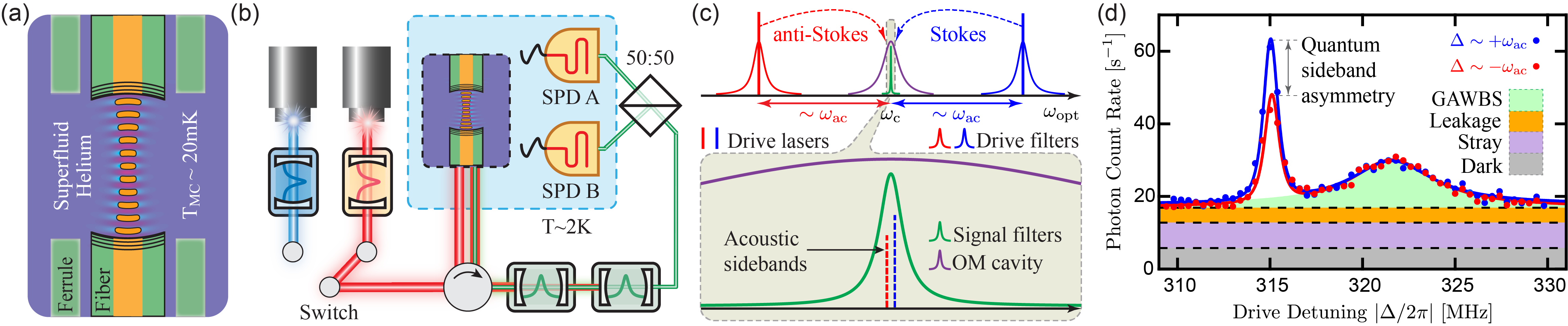}
\caption{\textbf{(a)} Device schematic: A fiber-based Fabry-Perot cavity is filled with superfluid $^4$He. Blue shading denotes the instantaneous $^4$He density in an acoustic mode. Orange denotes the optical mode intensity.
\textbf{(b)} Optical schematic showing the two drive lasers (red and blue paths), optomechanical cavity (OMC, black dashed box), acoustically scattered photons (green path), two signal filter cavities (green) and the two SPDs.
The filter cavities (red and blue) before the OMC are used to suppress laser phase noise.
\textbf{(c)} Optical spectrum showing the frequencies of the lasers, scattered photons, and filters, all with respect to the OMC's optical resonance.
\textbf{(d)} Photon count rate spectrum measured as a function of the drive laser detuning $\Delta$, with $P_\text{in} = 400$ nW.
}
\label{Fig1}
\end{figure*}

In this work, single photon detection is used to probe and control (via post-selection) the mechanical state of a $\sim 1$  ng oscillator comprised of superfluid $^4\text{He}$.
The oscillator's phonon coherences are measured up to the fourth order, and are found to be consistent with the acoustic mode having a Markovian coupling to its bath.
The phonon coherences of $k$-phonon-subtracted (and $k$-phonon-added) thermal states are also measured for $k \leq 3$.
These results provide a detailed characterization of the acoustic mode's environment, and demonstrate that superfluid mechanical elements are well-suited for accessing nonlinear quantum optomechanical effects at the nanogram scale. Several factors contribute to these devices' performance, including their simple geometry, the unique material properties of superfluid $^4$He, and the wide applicability of SPD-induced backaction \cite{delorenzo2017, shkarin2019}.

Figure ~\ref{Fig1}(a) shows a schematic of the device used here (also described in detail in Ref.~\cite{shkarin2019}).
Two single-mode optical fibers with high-reflectivity mirrors fabricated on their end faces are aligned using glass ferrules to form a Fabry-Perot optical cavity.
The ferrules and fibers are epoxied to a copper housing that is thermally anchored to the mixing chamber (MC) of a dilution refrigerator at temperature $T_{\mathrm{MC}}\approx 20$ mK, and the cavity is filled with superfluid $^4\text{He}$ via a capillary line. 
The fiber mirrors set equivalent boundary conditions for the cavity's optical and acoustic modes (the latter are density waves in the $^4\text{He}$); as a result, these modes' spatial profiles are well-approximated by a common set of orthogonal functions (the well-known Gaussian modes of paraxial cavities).
Since the optomechanical coupling is set by the overlap of the superfluid density fluctuations with the optical intensity, the orthogonality of these modes' spatial profiles ensures an unusually clean realization of single-mode optomechanics: a given optical mode with wavelength $\lambda_{\text{c}}$ (in $^4$He) couples only to the acoustic mode with wavelength $\lambda_{\text{ac}}=\lambda_{\text{c}}/2$.

When the optical mode is driven by a laser, the single-mode optomechanical interaction is described by the linearized Hamiltonian $\mathcal{H}_{\text{OM}} = - \hbar g_0 \sqrt{n_{\text{c}}} (a+a^\dagger) (b+b^\dagger)$, where $a$ and $b$ are the annihilation operators of the optical mode and of the acoustic mode respectively, $n_{\text{c}}$ is the mean photon number in the cavity, and $g_0$ is the single photon optomechanical coupling rate \cite{aspelmeyer2014}.

A schematic of the experiment is shown in Fig.~\ref{Fig1}(b,c).
The optomechanical cavity has an optical resonance at $\omega_{\text{c}}/2\pi=c/(n_\text{He}\lambda_\text{c})$ (corresponding to a vacuum wavelength $n_\text{He}\lambda_\text{c} = 1548.3(1)$ nm) and a linewidth $\kappa_\text{c}/2\pi = 47.2(5)$ MHz, where $n_\text{He}=1.0261$ is the refractive index of $^4\text{He}$.
It is driven with a laser which is either red-detuned from $\omega_\text{c}$ by $\Delta\sim-\omega_\text{ac}$, or else blue-detuned by $\Delta\sim+\omega_\text{ac}$, where $\omega_\text{ac}/2\pi=\nu_\text{He}/\lambda_\text{ac}$ is the acoustic mode frequency and $\nu_\text{He}=238$ m/s is the speed of sound in $^4\text{He}$.
The red- (blue-) detuned drive effectively realizes a beamsplitter (two-mode squeezing) optomechanical interaction via cavity-enhanced anti-Stokes (Stokes) scattering \cite{aspelmeyer2014}. 
Photons leaving the cavity (both the unshifted drive photons and the resonant anti-Stokes/Stokes photons) are then incident on two cavities which are arranged in series and have linewidths $\kappa_{\text{\tiny FC1}}/2\pi=1.71(2)\text{ MHz},\ \kappa_{\text{\tiny FC2}}/2\pi=1.21(5)\text{ MHz}$. These cavities' resonances are locked to $\omega_{\text{c}}$ \cite{seeSI}. Since they meet the condition $\gamma_{\mathrm{ac}} \ll \kappa_{\text{\tiny FC1,2}}$ (where the acoustic mode's linewidth $\gamma_{\mathrm{ac}}/2\pi \approx 3.5$ kHz) they serve as filters by reflecting the drive photons while passing the anti-Stokes/Stokes photons to superconducting nanowire SPDs. 

Figure \ref{Fig1}(d) shows a typical measurement of the photon detection rate as a function of $\Delta$.
The peaks at $\Delta/2\pi=\mp\omega_\text{ac}/2\pi=\mp315.3(1)$ MHz correspond to the anti-Stokes (Stokes) sidebands of the acoustic mode.
This frequency is consistent with the expected $\omega_\text{ac} = 315.40(2)$ MHz for the optical resonance employed ($\lambda_\text{ac}=\lambda_{\text{c}}/2=754.46(5)$ nm).
The broad peak at $\Delta/2\pi=\mp322.3(1)$ MHz is caused by guided acoustic wave Brillouin scattering (GAWBS) of drive laser photons in the room temperature optical fibers \cite{shelby1985}.
A detuning-independent background is also evident.
The solid lines in Fig.~\ref{Fig1}(d) are a fit to the sum of a constant (corresponding to the background counts), a broad Lorentzian (corresponding to the GAWBS signal), and the filter cavities' passband (a product of two Lorentzians, corresponding to the counts from the acoustic sidebands).
A detailed description of this fit is given in Ref.~\cite{seeSI}. 

\setlength{\abovecaptionskip}{0pt plus 1pt minus 1pt} 
\begin{figure*}
\centering
\includegraphics[width=0.98\textwidth]{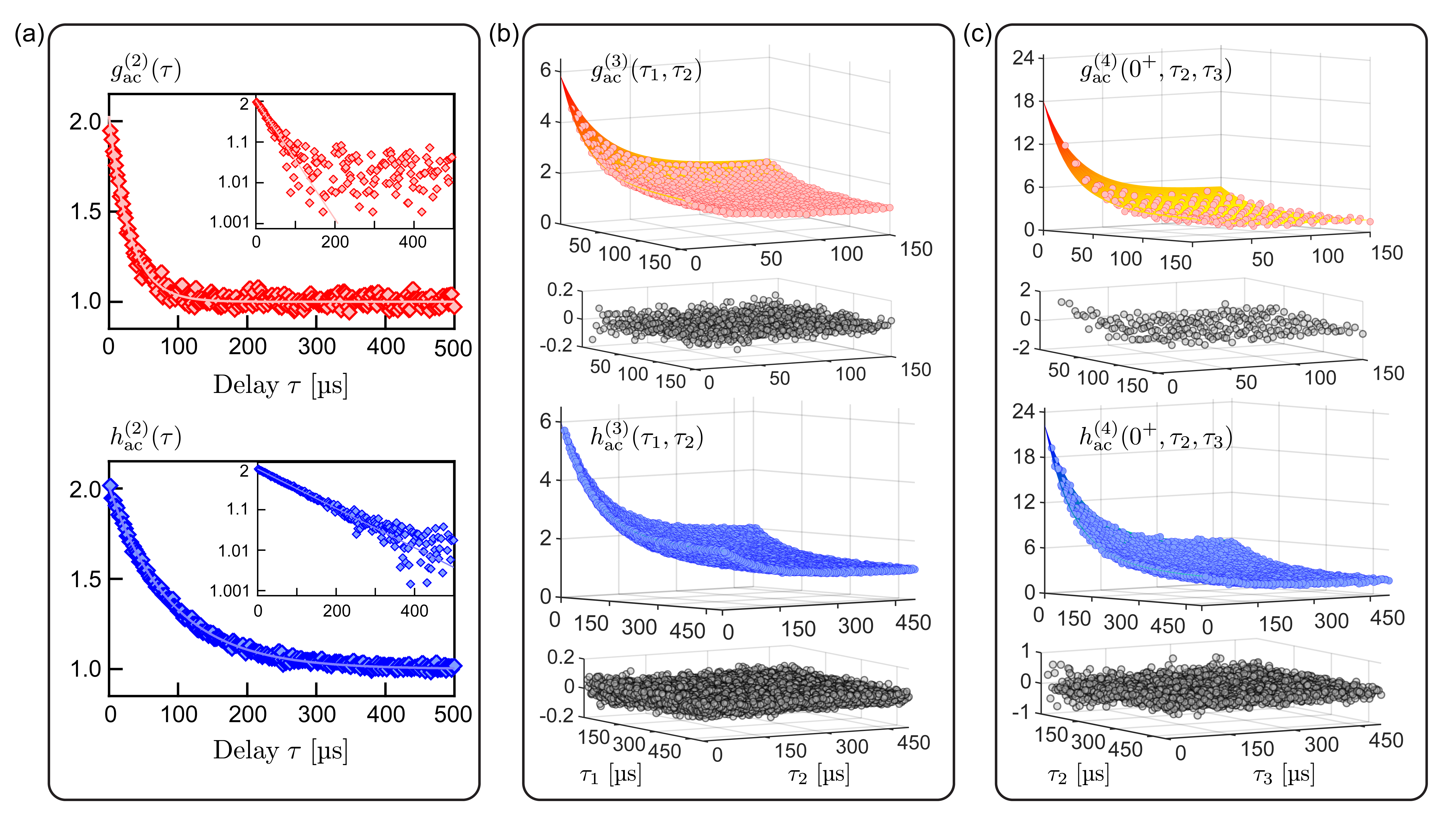}
\caption{Phonon coherences: \textbf{(a)} The second-, \textbf{(b)} third-, and \textbf{(c)} fourth- order phonon coherences measured for $P_\text{in}\approx$ \SI{5}{\micro\watt}, with photon arrival times binned in \SI{2}{\micro\second}, \SI{5}{\micro\second} and \SI{10}{\micro\second} bins respectively. In \textbf{(a)}, the insets show the same data on a logarithmic scale.
For the three-time dependent $g^{(4)}_\text{ac}(\tau_1,\tau_2,\tau_3)$ and $h^{(4)}_\text{ac}(\tau_1,\tau_2,\tau_3)$, we only show representative 2D slices of $g^{(4)}_\text{ac}(0^+,\tau_2,\tau_3)$ and $h^{(4)}_\text{ac}(0^+,\tau_2,\tau_3)$, where $\tau=0^+$ represents the bin with $\SI{5}{\micro\second}<\tau<\SI{15}{\micro\second}$.
See Ref.~\cite{seeSI} for other 2D slices.
Solid lines/surfaces show the fits described in the text.
Fits for (c) are to the entire 3D (i.e., $\tau_1$-, $\tau_2$-, $\tau_3$- dependent) data set.
Fit residuals are shown in black for (b) and (c).
}
\label{Fig3}
\end{figure*}

Fits as in Fig.\ \ref{Fig1}(d) yield the optomechanical scattering rates $R_\text{AS(S)} = \gamma_{\text{AS(S)}} \times \eta_\kappa \times \eta_{\text{det}}$, where $\gamma_{\text{AS(S)}}$ is the anti-Stokes (Stokes) scattering rate for $\Delta=-\omega_\text{ac}$ ($\Delta=+\omega_\text{ac}$), $\eta_\kappa=\kappa_{\mathrm{in}}/\kappa_{\mathrm{c}}$ is the cavity coupling efficiency, $\kappa_\text{in}$ is the cavity's coupling rate, and $\eta_{\text{det}}$ is the detection efficiency (set by the transmission of the filter cavities and the beam path, and by the SPD quantum efficiency).
Standard quantum optomechanics theory predicts that $\gamma_{\text{AS}} = \gamma_\text{ac} C n_\text{ac}$ and $\gamma_{\text{S}} = \gamma_\text{ac} C (n_\text{ac}+1)$, where $\gamma_\text{ac}$ is the `bare' acoustic damping rate, $C = 4 \frac{g_0^2}{\kappa_\text{c}\gamma_\text{ac}} n_\text{c}$ is the multi-photon cooperativity \cite{aspelmeyer2014}, and $n_\text{ac}=\langle b^\dagger b \rangle$.
The difference between $R_\text{AS} \propto n_\text{ac}$ and $R_\text{S}\propto (n_\text{ac}+1)$ is known as the quantum sideband asymmetry (QSA).

As shown in Ref.\ \cite{seeSI}, measurements of $R_\text{AS}$ and $R_\text{S}$ indicate that the acoustic mode's temperature $ T \approx T_\text{MC}$ when the incident laser power $P_\text{in} \lesssim 300$ nW. Measurements with $P_\text{in} > 300$ nW show the standard optomechanical damping effect, as well as heating (due to absorption of photons in the fibers and mirror coatings) that is consistent with a simple thermal model of the device. 

Measurements of the mean photon flux (as in Fig.~\ref{Fig1}(d)) provide information that could also be obtained by heterodyne measurements of the acoustic sidebands \cite{aspelmeyer2014}. However, much richer information is contained in the photon arrival times registered by the SPDs. This is because each detection of an anti-Stokes (Stokes) photon corresponds to the subtraction  (addition) of a phonon in the acoustic mode. For example, the coherence of anti-Stokes photons $g_\text{AS}^{(n)} = \langle (a_\text{AS}^\dagger)^n a_\text{AS}^n \rangle / \langle a_\text{AS}^\dagger a_\text{AS} \rangle ^n $ is equal to the normally ordered phonon coherence $g_\text{ac}^{(n)} \equiv \langle (b^\dagger)^n b^n \rangle / \langle b^\dagger b \rangle ^n$, while the coherence of Stokes photons  $g_\text{S}^{(n)} = \langle (a_\text{S}^\dagger)^n a_\text{S}^n \rangle / \langle a_\text{S}^\dagger a_\text{S} \rangle ^n $ is equal to the anti-normally ordered phonon coherence $h_\text{ac}^{(n)} \equiv \langle b^n (b^\dagger)^n \rangle / \langle b b^\dagger \rangle ^n$ \cite{seeSI}. Here $a_\text{AS}$ and $a_\text{S}$ are the annihilation operators for anti-Stokes and Stokes photons, respectively. 

Measurements of these phonon coherences can be used to probe the acoustic oscillator's dynamics. 
For example, an oscillator in a thermal state should exhibit phonon bunching that decays on a time scale set by the oscillator's damping.

If the coupling to the bath is Markovian, then the $n^{\text{th}} $-order coherence is predicted to be  $g^{(n)}_\text{ac}(\boldsymbol{\tau})=1 + f_n(\bar\gamma_\text{ac}\boldsymbol{\tau})$, where $\boldsymbol{\tau}=(\tau_1,...,\tau_{n-1})$, $\tau_k$ is the delay between the $k^\text{th}$ and $(k+1)^\text{th}$ detected phonon, and the oscillator's total damping rate is $\bar\gamma_\text{ac}(P_\text{in}) = \gamma_\text{ac}+\gamma_\text{opt}(P_\text{in})$, where $\gamma_\text{opt}(P_\text{in})$ is the contribution from optomechanical backaction \cite{seeSI}. The functions $f_n$ are straightforward to calculate, with $f_2(x) = e^{-x}$ and $f_3(\mathbf{x}) = e^{-x_1} + e^{-x_2} + 3e^{-x_1-x_2}$ (an expression for $f_4(\mathbf{x})$ is given in Ref. \cite{seeSI}).

To measure the optical coherences (and thus the phonon coherences), a histogram of the delays between $n$ photon arrival times $C_\text{AS(S)}^{(n)}(\boldsymbol{\tau})$ is constructed and then normalized by its value at large delays. 
In the experiment, the photon arrivals registered by the SPDs include the sideband photons as well as other events (such as background photons and dark counts, see Fig.~\ref{Fig1}(d)).
These extraneous events are measured to be independent and identically distributed over time, so their contribution to $C_\text{AS(S)}^{(n)}(\boldsymbol{\tau})$ can be calculated and corrected for \cite{seeSI}.
The corrected histograms  
are fit to the form $A+B\times f_n(\bar\gamma_\text{ac}\boldsymbol{\tau})$, where $A,B$, and $\bar\gamma_\text{ac}$ are fit parameters. 
The best-fit value of $A$ is used to normalize $C_\text{AS(S)}^{(n)}(\boldsymbol{\tau})$ and convert it to the corresponding phonon coherence (i.e., $g^{(n)}_\text{ac}(\boldsymbol{\tau})=C_\text{AS}^{(n)}(\boldsymbol{\tau})/A$ and $h^{(n)}_\text{ac}(\boldsymbol{\tau})=C_\text{S}^{(n)}(\boldsymbol{\tau})/A$).

\setlength{\abovecaptionskip}{0pt plus 1pt minus 1pt}
\begin{figure}[]
\centering
\includegraphics[width=0.48\textwidth]{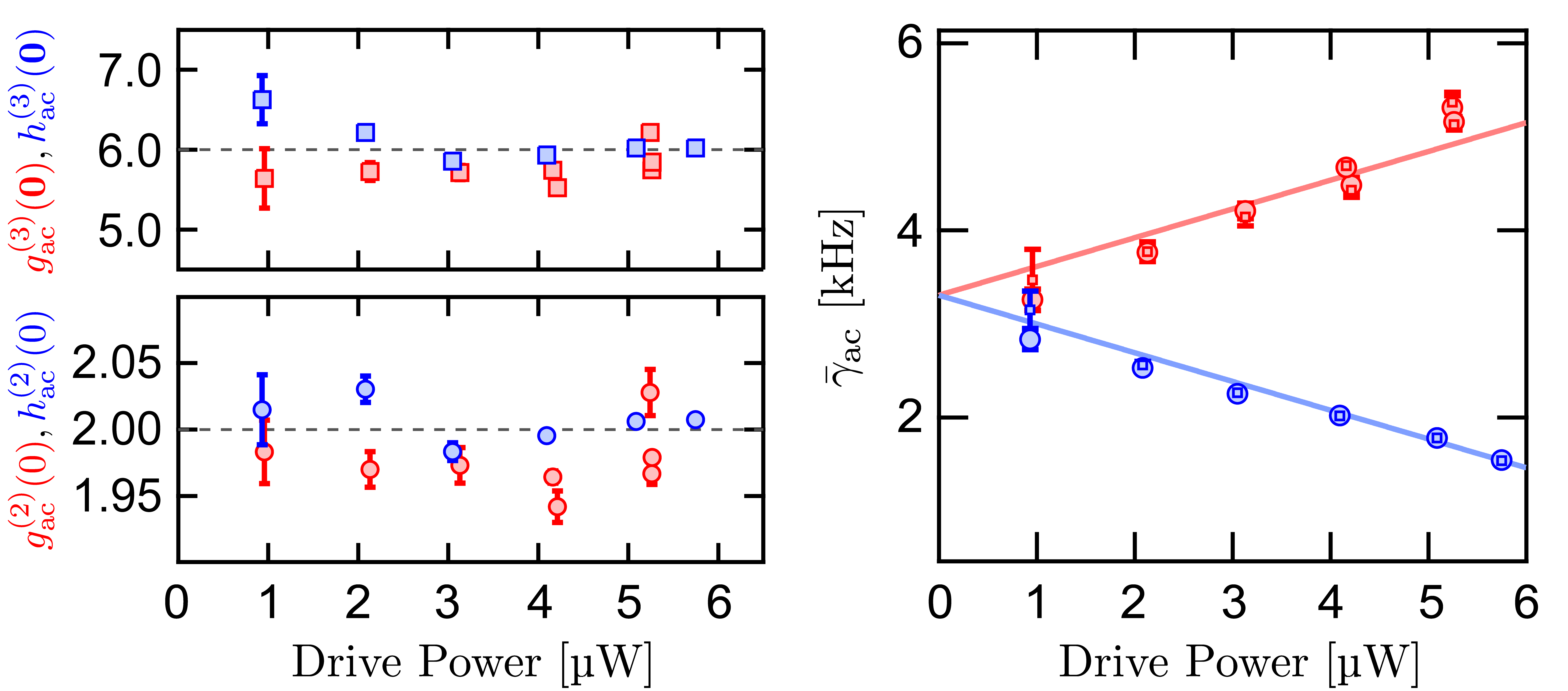}
\caption{The zero-delay second- and third- order coherences, and the coherence decay rates ($\bar\gamma_\text{ac}$), as a function of incident power $P_\text{in}$.
Data is extracted from fits to the second-order (circles) and third-order (squares) coherences.
Solid lines show a fit to standard optomechanics theory.
}
\label{Fig5}
\end{figure}

Figure \ref{Fig3} shows the phonon coherences measured in this way (up to the fourth order) as a function of delay times, along with the corresponding fits.
The zero-delay coherence values extracted from these fits are $g^{(2)}_{\mathrm{ac}}(0)=1.980(2)$, $h^{(2)}_{\mathrm{ac}}(0)=2.007(1)$, $g^{(3)}_{\mathrm{ac}}(\mathbf{0})=5.843(7)$, $h^{(3)}_{\mathrm{ac}}(\mathbf{0})=6.023(2)$, $g^{(4)}_{\mathrm{ac}}(\mathbf{0})=23.01(3)$, and $h^{(4)}_{\mathrm{ac}}(\mathbf{0})=23.98(1)$ (where the stated uncertainty corresponds to one standard deviation of the best-fit parameter).  These values are consistent with the predictions for a thermal state: $g^{(n)}_{\mathrm{ac}}(\mathbf{0}) = h^{(n)}_{\mathrm{ac}}(\mathbf{0}) = n!$.
(The fourth-order data and fits shown in Fig.~\ref{Fig3}(c) are for a finite delay bin of $\SI{5}{\micro\second}<\tau_1 <\SI{15}{\micro\second}$, and are thus expected to be less than $4!=24$ for $(\tau_2,\tau_3)\rightarrow (0,0)$.)
The $\boldsymbol{\tau}$-dependence of the coherences also agrees well with theory, as evidenced by the small residuals.
This demonstrates that the acoustic mode is in equilibrium with the bath and that its energy fluctuations are consistent with a Gaussian distribution (to at least the fourth cumulant).

Figure \ref{Fig5} shows various features of these fits for \SI{0.9}{\micro\watt} $ < P_\text{in} <$ \SI{6}{\micro\watt} (corresponding to $1  \lesssim n_\text{ac} \lesssim 10$ \cite{seeSI}). The left panel shows that the zero-delay coherences $g^{(2)}_{\mathrm{ac}}(0)$ and $g^{(3)}_{\mathrm{ac}}(\mathbf{0})$ are close to $2$ and $6$, respectively, for all $P_\text{in}$ in this range. The right panel shows that the decay rates $\bar\gamma_\text{ac}(P_\text{in})$ extracted from fits (as in Fig.~\ref{Fig3}) exhibit the expected optomechanical backaction. A fit to standard optomechanics theory \cite{aspelmeyer2014} (solid lines) gives  $g_0/2\pi=4.70(5)$ kHz, consistent with the independent calibration described in Ref. \cite{seeSI}.

\setlength{\abovecaptionskip}{0pt plus 1pt minus 1pt}
\begin{figure}[]
\centering
\includegraphics[width=0.48\textwidth]{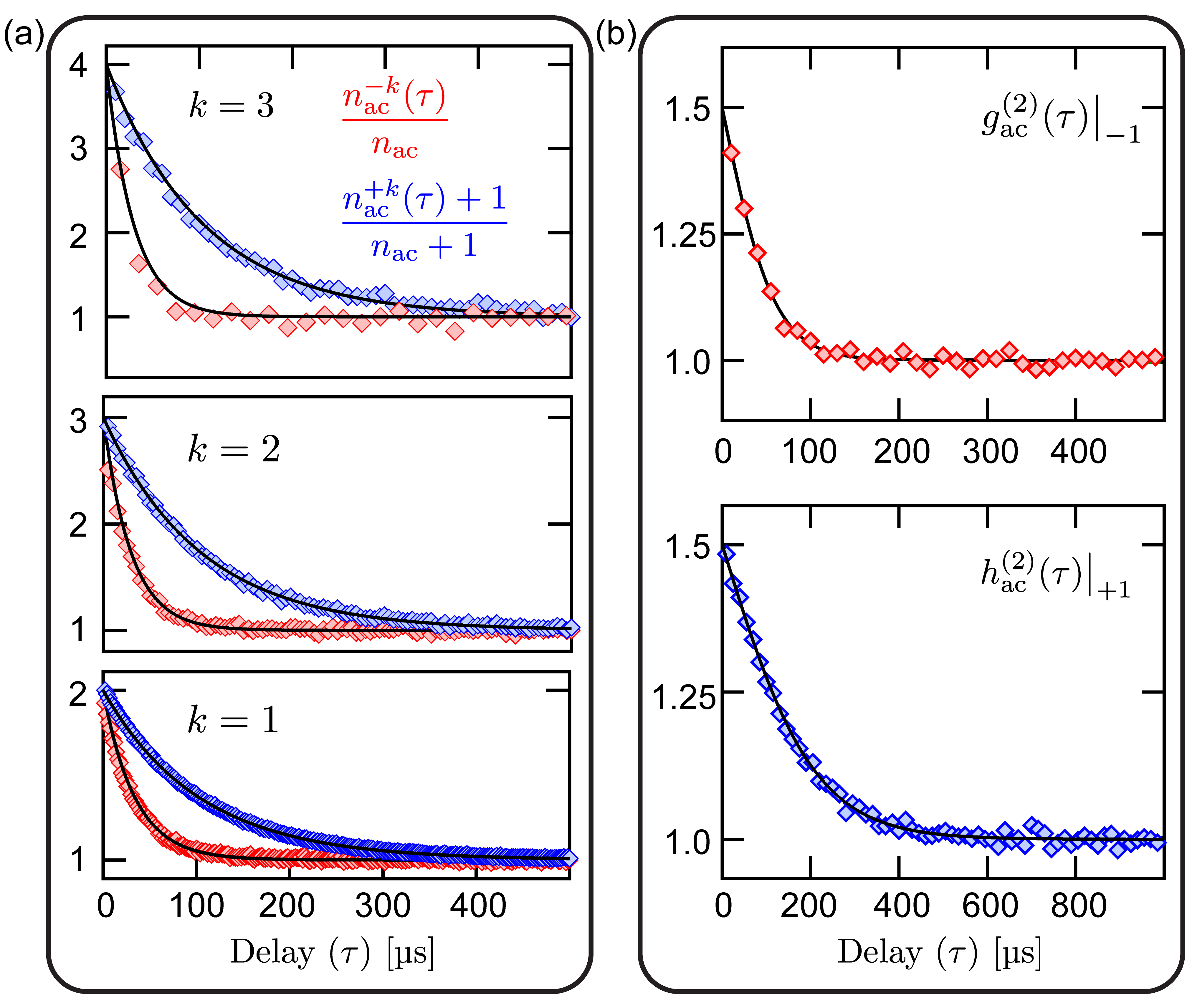}
\caption{
\textbf{(a)} Dynamics of the mean phonon occupancy upon subtraction/addition of $k$ phonons at $\tau=0$.
\textbf{(b)} Second order coherences of a 1-phonon subtracted (red) and added (blue) thermal state. 
Solid lines show the theoretical predictions, see Ref.~\cite{seeSI}.
Data shown for $P_\text{in}\approx$ \SI{5}{\micro\watt}.
}
\label{Fig4}
\end{figure}

The analysis described above (and shown in Fig.\ \ref{Fig3} and Fig.\ \ref{Fig5}) utilizes all the photons registered by the SPD.
However, by post-selecting data that is recorded immediately after detection of one or more anti-Stokes (Stokes) photons, one can measure the properties of phonon-subtracted (phonon-added) states.
For instance, $g^{(2)}_\text{ac}(0)=2$ implies that the mean rate of photon arrivals doubles immediately after the detection of one anti-Stokes photon (or equivalently, the subtraction of a phonon).
As the scattering rate $\gamma_\text{AS}$ is proportional to the acoustic mode's mean phonon occupancy $n_\text{ac}$, one can conclude that $n_\text{ac}$ doubles after the subtraction of a phonon.
More generally, the evolution of the mean phonon occupancy $n^{-k}_\text{ac}(\tau)$ ($n^{+k}_\text{ac}(\tau)$) of a $k-$phonon subtracted (added) state can  be measured through appropriate post-selection. See Ref. \cite{seeSI} for details.

Measurements for $k=1,2,3$ are shown in Fig.~\ref{Fig4}(a). 
If the equilibrium state (i.e., the state before the $k$-phonon subtraction/addition event) is thermal, $n_\text{ac}^{-k}(0) = (k+1) n_\text{ac}$, i.e. the mean occupancy increases $(k+1)$-fold on the subtraction of $k$ phonons, while $n_\text{ac}^{+k}(0) = (k+1)n_\text{ac}+k$. 
As seen in Fig.~\ref{Fig4}(a), the phonon occupancy is indeed measured to double/triple/quadruple immediately after 1-/2-/3- phonon subtraction, and to subsequently decay back to equilibrium occupancy with the predicted time dependence (solid lines).

Similarly, by appropriately post-selecting and analyzing the recorded photon arrivals, it is possible to construct the various coherences of $k$-phonon subtracted (or added) thermal states \cite{seeSI}. Fig.~\ref{Fig4}(b) shows the measured second-order coherence of a $1$-phonon subtracted thermal state (normally ordered $g^{(2)}_{\mathrm{ac}}(\tau)|_{-1}$), and of a $1$-phonon added thermal state (anti-normally ordered $h^{(2)}_{\mathrm{ac}}(\tau)|_{+1}$), along with their theoretical expectations (solid lines). 
The measured zero-time second order coherences agree well with the theoretical expectation of $3/2$, as does their decay to unity on the mechanical timescale.

Coherences and other statistics of $k$-quanta -subtracted/-added thermal states are of interest in quantum metrology, quantum information and quantum thermodynamics.
The optical equivalents of such states have been shown to be efficient at performing work and carrying information \cite{hlousek2017}.
The ability to create and probe these states in an acoustic mode, as demonstrated here, extends the potential use of such states to optomechanical platforms \cite{enzian2020, patel2021, enzian2021}.

We thank Radim Filip, Sebastian Garcia, Chitres Guria, Steve Girvin, Anna Kashkanova, Konstantin Ott, Andrey Rakhubovsky, and Alexey Shkarin for their help. This work is supported by the NSF (Award No.~1707703), AFOSR (Grant No.~FA9550-15-1-0270), the Vannevar Bush Faculty Fellowship (No.~N00014-20-1-2628), and by the Quantum Information Science Enabled Discovery (QuantISED) for High Energy Physics (KA2401032). KJ acknowledges support from the IC Postdoctoral Research Fellowship. 

\bibliography{ms}

\end{document}


\author{Y. S. S. Patil}
\email{yogesh.patil@yale.edu}
\affiliation{Department of Physics, Yale University, New Haven, Connecticut 06520, USA}
\author{J. Yu}
\affiliation{Department of Physics, Yale University, New Haven, Connecticut 06520, USA}
\author{S. Frazier}
\affiliation{Department of Physics, Yale University, New Haven, Connecticut 06520, USA}
\author{Y. Wang}
\affiliation{Department of Applied Physics, Yale University, New Haven, Connecticut 06520, USA}
\author{K. Johnson}
\affiliation{Department of Physics, Yale University, New Haven, Connecticut 06520, USA}
\author{J. Fox}
\affiliation{Department of Physics, Yale University, New Haven, Connecticut 06520, USA}
\author{J. Reichel}
\affiliation{Laboratoire Kastler Brossel, ENS-Universit\'{e} PSL, CNRS, Sorbonne Universit\'{e}, Coll\`{e}ge de France 24 rue Lhomond, 75005 Paris, France}
\author{J. G. E. Harris}
\email{jack.harris@yale.edu}
\affiliation{Department of Physics, Yale University, New Haven, Connecticut 06520, USA}
\affiliation{Department of Applied Physics, Yale University, New Haven, Connecticut 06520, USA}
\affiliation{Yale Quantum Institute, Yale University, New Haven, Connecticut 06520, USA}

\title{Supplemental Information: \texorpdfstring{\\}M 
Measuring High-Order Phonon Correlations in an Optomechanical Resonator}
\maketitle

\setlength\abovedisplayskip{3pt}
\setlength\belowdisplayskip{3pt}

\section*{}

\subsection{Optical Setup}
\label{opticalSetupSection}
This section describes the optical setup used for the experiments presented in the main text. 

\begin{figure}
\centering
\includegraphics[width=0.48\textwidth]{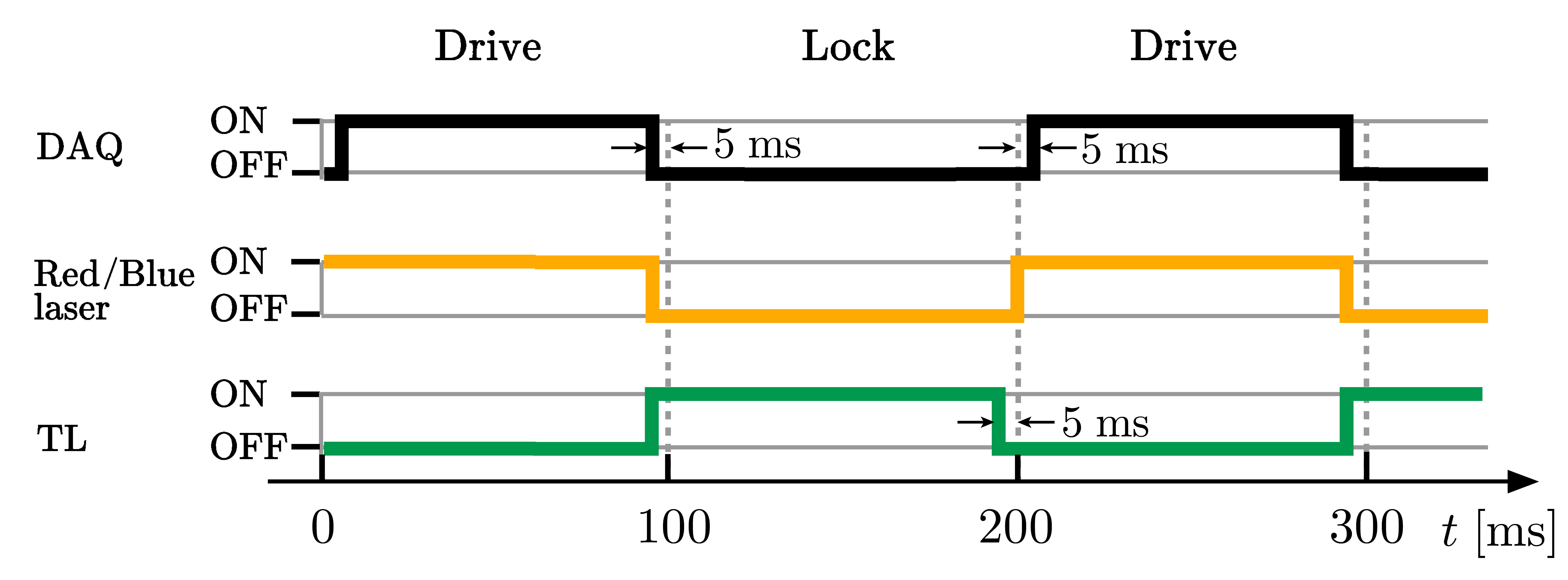}
\caption{
\textbf{Lock and drive sequence.} The time sequence for switching between the ``lock'' and the ``drive'' configurations. Photon counts are acquired only during the central 90 ms of the ``drive'' period.
}
\label{LHSequence}
\end{figure}

\begin{figure*}[ht]
\centering
\begin{adjustbox}{addcode={\begin{minipage}{\width}}
{\caption{
    \textbf{Optical schematic.} This figure shows the two optical configurations (lock and drive) used in the experiment. The tunable laser (TL) is locked to the optomechanical cavity during the lock configuration. The blue (BL) and red (RL) drive lasers are always locked to the TL with a frequency offset $\sim \pm\omega_\text{ac}$. A total of six $1 \times 2$ MEMS switches are used to switch between the lock configuration (green optical path) and the drive configuration (orange optical path) every 100 ms, such that the optical components are either locked to their desired frequency, or free running (as indicated in Fig.~\ref{LHSequence}). PC: polarization controller. VOA: variable optical attenutator. EDFA: erbium-doped fiber amplifier. EOM: electro-optic modulator. ILP: in-line polarizer. ISO: isolator. FC: filter cavity.
    }\label{opticalSetup}
    \end{minipage}},rotate=90,center}
    \includegraphics[scale=0.8,angle=0]{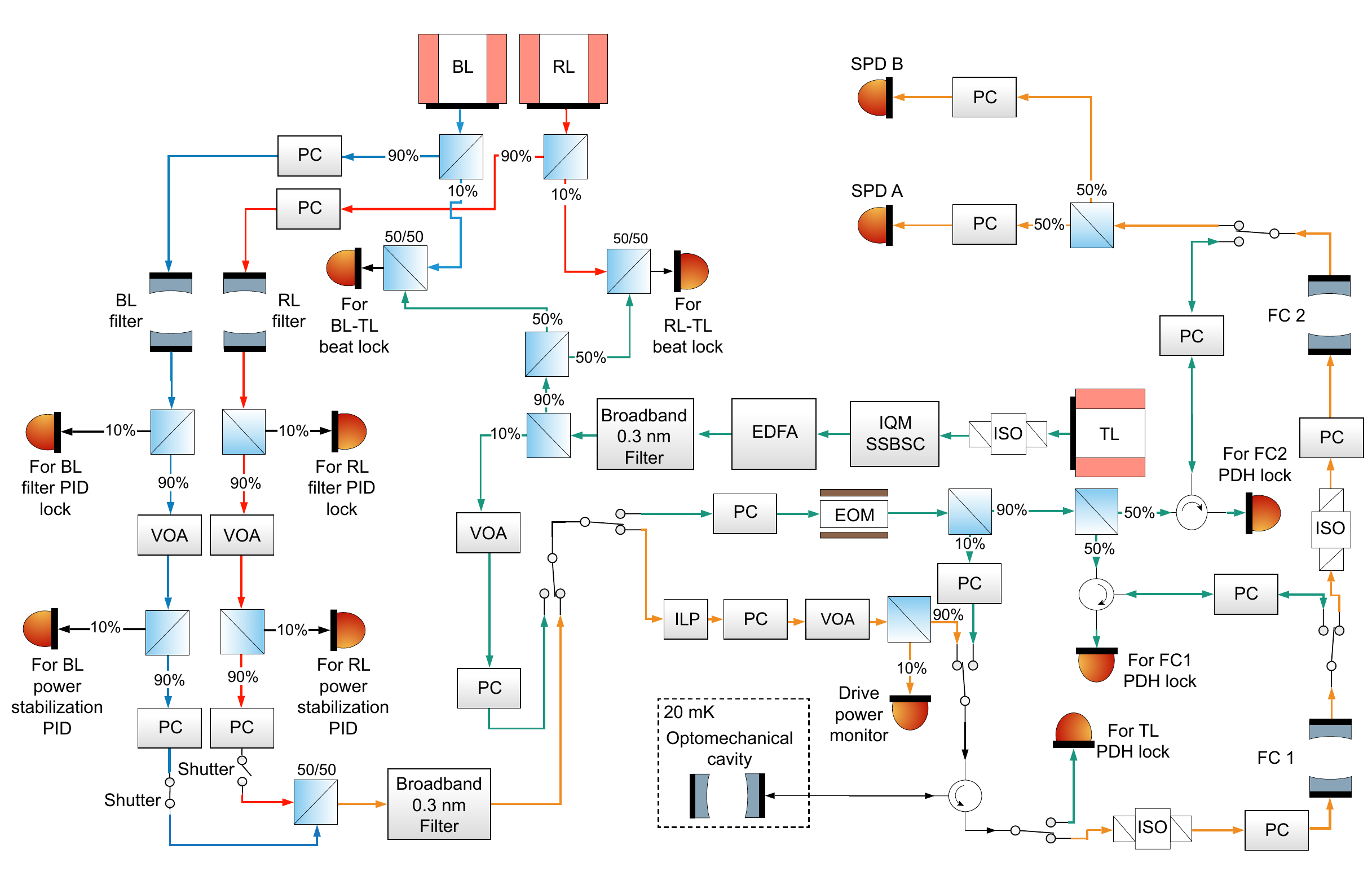}
\end{adjustbox}
\end{figure*}

Before describing the setup itself, we first describe the timing sequence used in these measurements. During an experiment the setup is alternated between two configurations, each lasting 100 ms. The time sequence is shown in Fig.~\ref{LHSequence} (the configurations are shown in Fig.~\ref{opticalSetup} and discussed in the following paragraph). The first of these is the ``lock" configuration, during which the optical path is configured to lock the tunable laser (TL) and filter cavities to the optomechanical cavity (OMC). The second is the ``drive" configuration, during which the various control voltages used to tune the TL and filter cavities are held constant. Also, at the start of the drive period, the optical path is configured so that either a blue- or red-detuned laser drives the OMC. Photon counting data is collected during the middle 90 ms of the drive period. 

Figure \ref{opticalSetup} shows the two configurations, with the lock configuration's optical path in green and the drive configuration's optical path in orange. The lock configuration begins with the TL entering an IQ modulator (IQM) operating in the single-sideband suppressed-carrier mode. The IQM serves as a frequency shifter to lock the TL to the OMC. The output light from the IQM is amplified by an Erbium-doped fiber amplifer (EDFA) and filtered by a broadband ($\Delta \lambda \approx$ 0.3 nm) filter to suppress the EDFA's amplified spontaneous emission noise. The beam then passes through an electro-optic modulator (EOM) which produces a pair of sidebands used for standard Pound-Drever-Hall (PDH) locking. The beam is then split into three parts, which are sent to the OMC and the two filter cavities (FC1, FC2). The TL is thus locked to the OMC via the IQM control voltage, and the filter cavities are locked to the TL via their piezo and thermal tuning.

\setlength{\abovecaptionskip}{0pt plus 2pt minus 2pt} 

\setlength{\abovecaptionskip}{0pt plus 2pt minus 2pt} 

The blue-/red-detuned drive lasers (BL/RL) are always locked to the TL. This is accomplished by combining each drive laser with the TL (shown in the upper left corner of Fig.~\ref{opticalSetup}) and monitoring the frequency of the resulting beat note to ensure that the two drive lasers are detuned from the OMC by $\sim$ $\pm\omega_\text{ac}$. Each drive laser passes through a filter cavity ($\kappa/2\pi \approx$ 30 MHz) and is then combined via a 50/50 fiber coupler to drive the OMC. The part of the optical path that is common to the two drive beams is shown in orange. A pair of shutters just before the 50/50 coupler are used to determine whether the BL or the RL drives the OMC. The shutters are also used to set the desired pulse sequence for pulsed measurements.

Scattered sideband photons that are resonant with the OMC exit the cavity and pass through filter cavities FC1 and FC2, finally arriving at a pair of superconducting nanowire single photon detectors (SNSPDs). The variable optical attenuators (VOAs) are used to adjust the optical power, and the polarization controllers (PCs) are used to align the polarization of light entering polarization-sensitive components such as the FCs, the EOM and the SNSPDs.

\subsection{Characterizating the Spectrum of Photon Counts}\label{background count}
\begin{figure}
\centering
\includegraphics[width=0.48\textwidth]{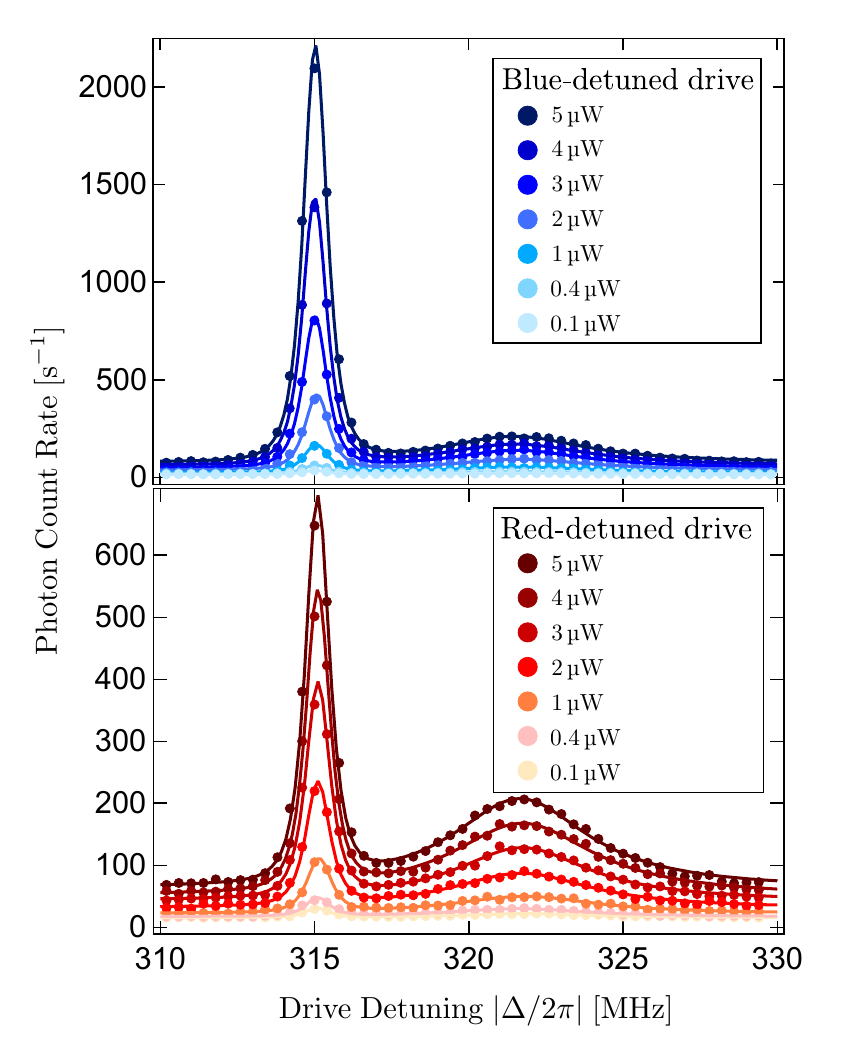}
\caption{
\textbf{Photon count spectrum.} Count rates as a function of the detuning and power of the drive laser. The narrow peak near $315$ MHz corresponds to sideband photons produced by the creation and annihilation of phonons in the superfluid-filled cavity. The broad beak near $322$ MHz is due to thermal fluctuations in the room-temperature optical fibers. The fits are described in the text.
}
\label{FigWideScan}
\end{figure}

Figure 1(d) of the main text shows a typical photon count rate spectrum. Here we discuss the characterization of such spectra. Fig.~\ref{FigWideScan} shows the count rate spectra for various values of the drive power $P_\text{in}$. Each spectrum shows a narrow peak (arising from the Stokes- and anti Stokes-scattered photons) centered at $\omega_\text{ac}/2\pi \approx$ 315.3 MHz, a broader peak centered around 322.3 MHz, and a frequency-independent background. 

\setlength{\abovecaptionskip}{0pt plus 2pt minus 2pt} 
\begin{figure}
\centering
\includegraphics[width=0.48\textwidth]{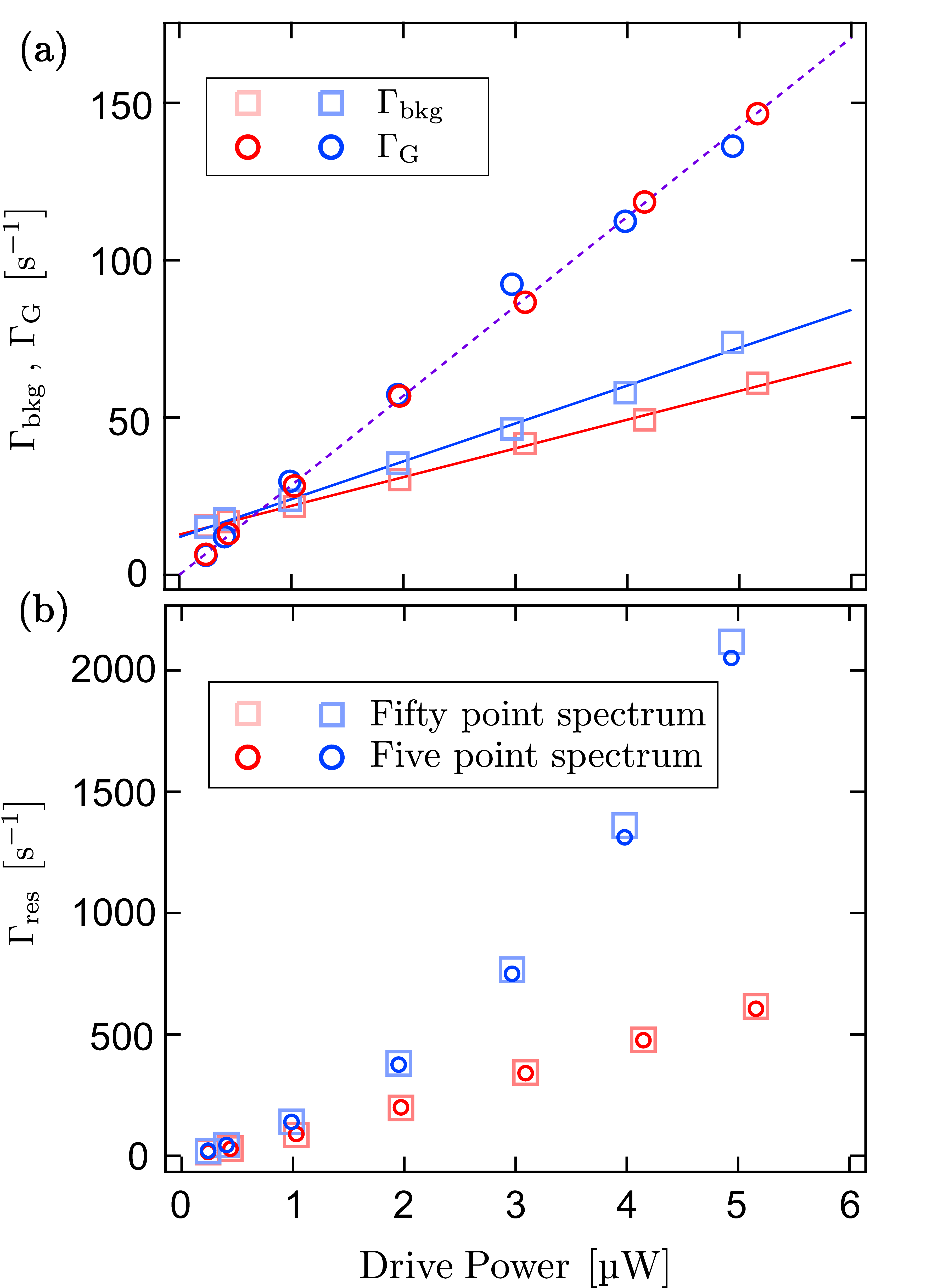}
\caption{
\textbf{Power dependence of the photon count spectrum. (a)} $\Gamma_\mathrm{bkg}$ and $\Gamma_\mathrm{G}$ as a function of drive power, and the corresponding linear fits. For $\Gamma_\text{bkg}$, the y-intercepts give background counts arising from SNSPD dark counts and stray light from the environment. 
The slopes give the leakage of drive photons around the filter cavities for the red-detuned and blue-detuned drive each. 
$\Gamma_\mathrm{G}$ also increases linearly with drive power.
\textbf{(b)} $\Gamma_\mathrm{res}$ as a function of drive power. Shown as open squares are values of $\Gamma_\mathrm{res}$ extracted from a 50-frequency measurement that is fit to Eq.~\ref{EqWideScanFit}. 
$\Gamma_\mathrm{res}$ extracted from a five-frequency measurement that is fit to Eq.~\ref{Eq5ptFit} are shown as open circles.
The difference between the two methods is negligible.
}
\label{spectrumFit}
\end{figure}

The sideband peak is expected to be proportional to the transfer function of the two cascaded filter cavities FC1 and FC2, given by 
\begin{align}
    f_\text{filter}(\Delta)=\frac{1}{
    \bigg[1+ 4 \big(\frac{|\Delta|-\omega_\text{ac}}{\kappa_{\mathrm{FC}1}}\big)^2\bigg]
    \bigg[1+ 4 \big(\frac{|\Delta|-\omega_\text{ac}}{\kappa_{\mathrm{FC}1}}\big)^2\bigg]
    }
\label{FilterFunc}
\end{align}
where $\kappa_{\mathrm{FC}1}/2\pi$ = 1.21(5) MHz and $\kappa_{\mathrm{FC}2}/2\pi$ = 1.71(2) MHz are the filter linewidths. We identify the broad peak around 322 MHz as the signature of guided-acoustic wave Brillouin scattering (GAWBS) in the fibers (See Sec.~\ref{GAWBSsec} for more details), which can be characterized by a Lorentzian centered at $\omega_\text{G}$:
\begin{align}
    f_{\text{G}}(\Delta) = \frac{1}{1 + 4 \big( \frac{|\Delta|-\omega_\text{G}}{\kappa_\text{G}} \big)^{2} }
\label{GAWBSFunc}
\end{align}

Thus, for each value of $P_\text{in}$ and for each detuning range (i.e., $\Delta \sim + \omega_{\text{ac}} $ and $\Delta \sim - \omega_{\text{ac}} $ ), the count rate spectrum is fit to the form 
\begin{align}
    \Gamma(\Delta) = \Gamma_\text{bkg} + f_\text{filter}(\Delta)\Gamma_\text{res} + f_\text{G}(\Delta)\Gamma_\text{G}
\label{EqWideScanFit}
\end{align}
where $\Gamma_\text{bkg}$, $\Gamma_\text{res}$, $\Gamma_\text{G}$, $\omega_\text{ac}$, $\omega_\text{G}$, and $\kappa_\text{G}$ are the fit parameters.

Fig.~\ref{spectrumFit} shows the fit-extracted $\Gamma_\text{bkg}$, $\Gamma_\text{G}$, and $\Gamma_\text{res}$ as a function of $P_\text{in}$. The frequency-independent background $\Gamma_\text{bkg}$ is found to have three sources: 1) the SNSPDs' darks counts, 2) stray light leaking into the fibers, and 3) unfiltered drive photons leaking around the two filter cavities in the detection chain (this last contribution is polarization-dependent, and so differs between the red-detuned and blue-detuned drives). 
$\Gamma_\text{bkg}(P_\text{in})$ is fit to the form $\Gamma_\text{bkg,0}$ + $\Gamma_\text{bkg,1} P_\text{in}$, which gives $\Gamma_\text{bkg,0}$ = 12.4 $\pm$ 0.8 s$^{-1}$, and $\Gamma_\text{bkg,1} = 9.1 \pm 0.2$ s$^{-1}$\textmu W$^{-1}$ for the red-detuned drive, and $\Gamma_\text{bkg,1} = 12.0 \pm 0.4$ s$^{-1}$\textmu W$^{-1}$ for the blue-detuned drive. By acquiring the count rate while blocking the input port directly in front of the SNSPDs, we 
find that $\Gamma_\text{bkg,0}$ consists of 7 $\pm$ 1 s$^{-1}$ from dark counts and 5 $\pm$ 1 s$^{-1}$ from stray light leaking into the detection-chain fibers. $\Gamma_\text{bkg,1}$ quantifies the amount of drive photons that leak around the filter cavities.

$\Gamma_\text{G}(P_\text{in})$ is fit to the form $\Gamma_\text{G,1} P_\text{in}$. This gives $\Gamma_\text{G,1}$ = 28.5 $\pm$ 0.3 s$^{-1}$\textmu W$^{-1}$.

A complete characterization of $\Gamma_\text{res}$ as a function of drive power is given in Sec.~\ref{drivePowerScanSec}. 

The spectra in Fig.~\ref{FigWideScan} (and those used to produce Fig.~\ref{spectrumFit}) are measured at $50$ values of $\Delta / 2 \pi$. While this is helpful in characterizing the device, the focus of the main paper is on the Stokes- and anti-Stokes- scattered photons. To measure the rates of these specific photons, we found that it was adequate to record the spectrum at just five frequencies close to $\omega_\text{ac}$ (specifically, $\Delta / 2 \pi =$ 310, 312, 314.9, 315.4, 315.9 MHz), and then fit the results to the form
\begin{align}
    \Gamma(\omega) = \Gamma_\text{bkg} + f_\text{filter}(\omega)\Gamma_\text{res}
\label{Eq5ptFit}
\end{align}

To illustrate the validity of this approach, Fig.~\ref{spectrumFit}(b) shows $\Gamma_\text{res}(P_\text{in})$ acquired in two ways: by measuring the count rate at $50$ values of $\Delta/ 2 \pi$ ranging from $\pm$ 310 MHz to $\pm$ 330 MHz (as in Fig.~\ref{FigWideScan}) and fitting to Eq.~\ref{EqWideScanFit}; and by measuring the count rate at just the five values of $\Delta/ 2 \pi$ listed above and fitting to Eq.~\ref{Eq5ptFit}. The difference is negligible, showing the reliability of the five-point spectrum and the minimal influence of the GAWBS peak (which is not included in Eq.~\ref{Eq5ptFit}) on this analysis. The data presented in Sec.~\ref{drivePowerScanSec}  were acquired using the ``five frequency'' approach.

\setlength{\abovecaptionskip}{0pt plus 2pt minus 2pt} 
\begin{figure}[]
\centering
\includegraphics[width=0.48\textwidth]{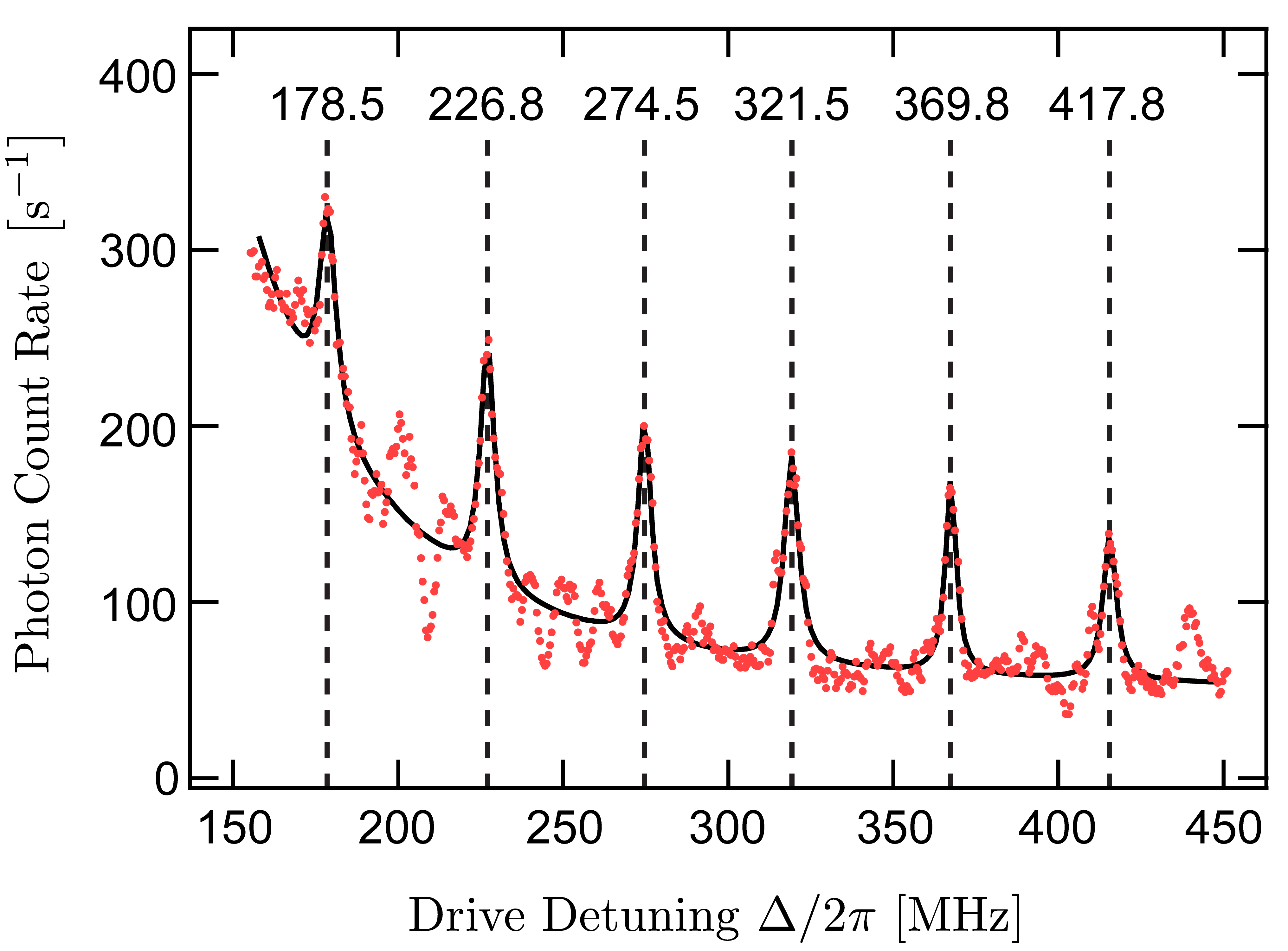}
\caption{
\textbf{Wide frequency scan showing several GAWBS peaks.} Six individual Lorentzians are fit (in addition to the tail of $f_{\mathrm{filter}}$), with the vertical dashed lines marking the measured (best-fit) center frequencies for the fiber's transverse acoustic modes.
}
\label{FigWiderScan}
\end{figure}

\setlength{\abovecaptionskip}{0pt plus 2pt minus 2pt} 
\begin{figure}
\centering
\includegraphics[width=0.48\textwidth]{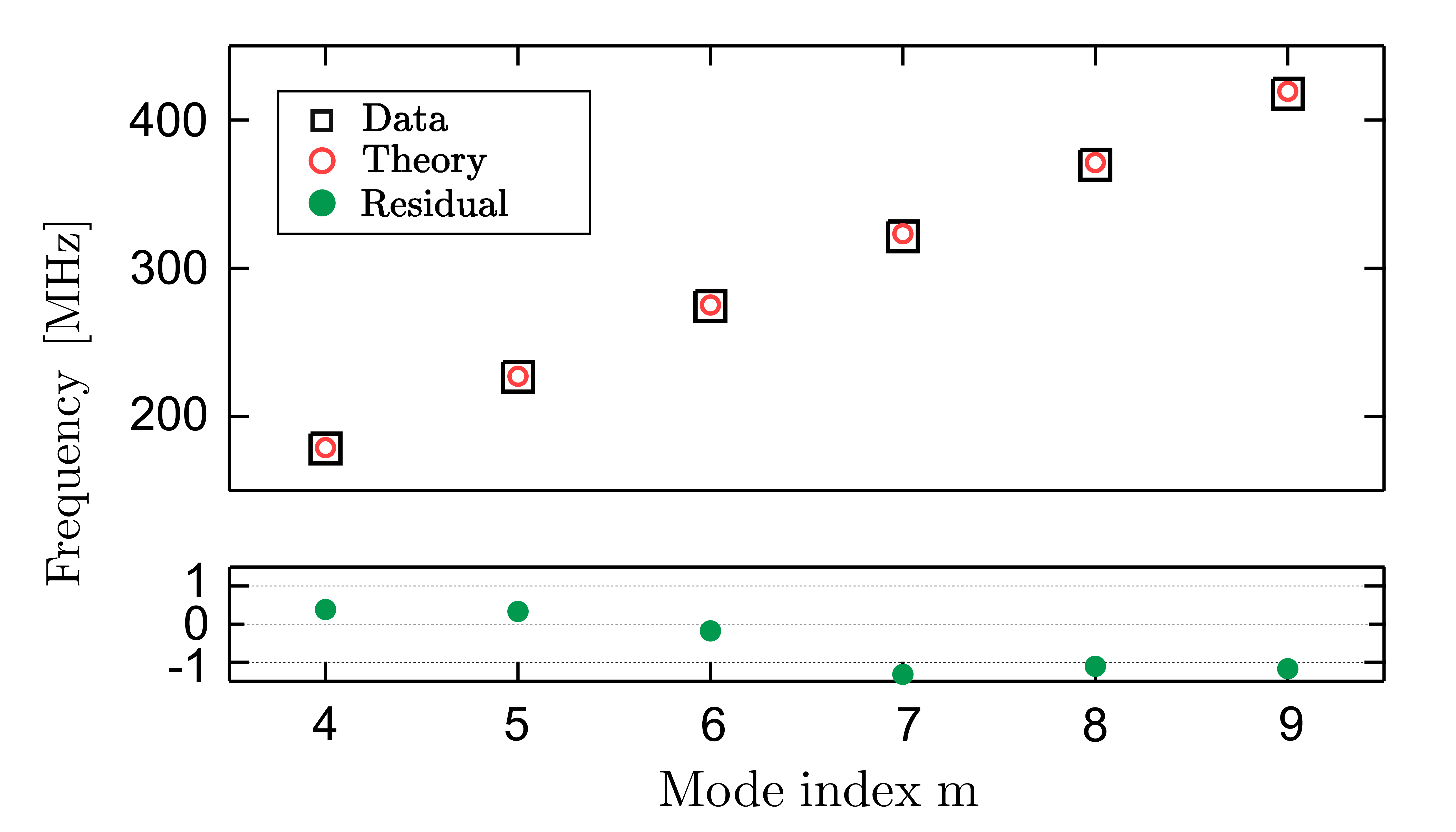}
\caption{
\textbf{The resonant frequencies of GAWBS.} The fitted peak frequencies are shown in solid black squares and the calculated frequencies of $R_{0m}$ GAWBS resonant modes for a \SI{125}{\micro\meter} single-mode fused silica fiber are shown in open red circles. The discrepancies are shown on the bottom graph in solid green circles.
}
\label{gawbsFreq}
\end{figure}

\subsection{Measuring the Wider Photon Count Spectrum and Characterizing GAWBS} \label{GAWBSsec}
This section gives a more detailed characterization of the GAWBS signature mentioned in Sec.~\ref{background count}. GAWBS is the well-studied interaction between light in a fiber and the fiber's thermally populated transverse acoustic modes \cite{shelby1985}. In practice, this interaction produces phase noise in the light with a spectrum determined by the fiber's acoustic modes.

Figure \ref{FigWiderScan} shows the photon count rate as the detuning $\Delta/ 2 \pi$ is varied from 150 to 450 MHz. The overall descending background in Fig.~\ref{FigWiderScan} corresponds to the tail of the filter cavity's transfer function. Due to the large steps in the detuning, the sideband peak of the acoustic cavity mode cannot be seen here, but several other peaks are evident. These peaks are found to be centered close to the frequencies $f_m$ that are expected for the fiber's transverse acoustic modes,  $R_{0m}$, which vibrate in the radial direction. These frequencies (labeled as vertical lines in Fig.~\ref{FigWiderScan}) correspond to the solutions $y_m$ of
\begin{align}
    (1-\alpha^{2})J_{0}(y) = \alpha^{2}J_{2}(y)
\label{EqYGawbs}
\end{align}
where $J_x$ is the $x$-order Bessel function of the first kind, and $\alpha = 0.624$ is the ratio between transverse and longitudinal phonon velocities for fibers made of fused silica.
The resonance frequencies are given by $f_m = \frac{V_d}{2\pi a}y_m$, where $V_\text{d} = 5996$ m/s is the longitudinal phonon velocity and the fiber radius $a$ = 62.5 \textmu m \cite{shelby1985, takefushi2019}.

The data in Fig.~\ref{FigWiderScan} is fit to the sum of six Lorentzians (whose center frequencies, linewidths, and magnitudes are the fit parameters), and the tail of $f_\text{filter}(\omega)$. The best-fit values for the center frequencies and the theoretical $f_m$ calculated from Eq.~\ref{EqYGawbs} are shown in  Fig.~\ref{gawbsFreq}.

\subsection{Model and Fits for Drive Power Scans}\label{drivePowerScanSec}
This section describes measurements of the sideband-photon counting rates as a function of the mixing chamber temperature $T_\text{MC}$ and the laser power incident on the device $P_\text{in}$. These measurements provide information about the acoustic mode's temperature and the useful range of $P_\text{in}$. They also provide additional calibration of the device's optomechanical coupling rate $g_0$.

\begin{figure}[]
\centering
\includegraphics[width=0.47\textwidth]{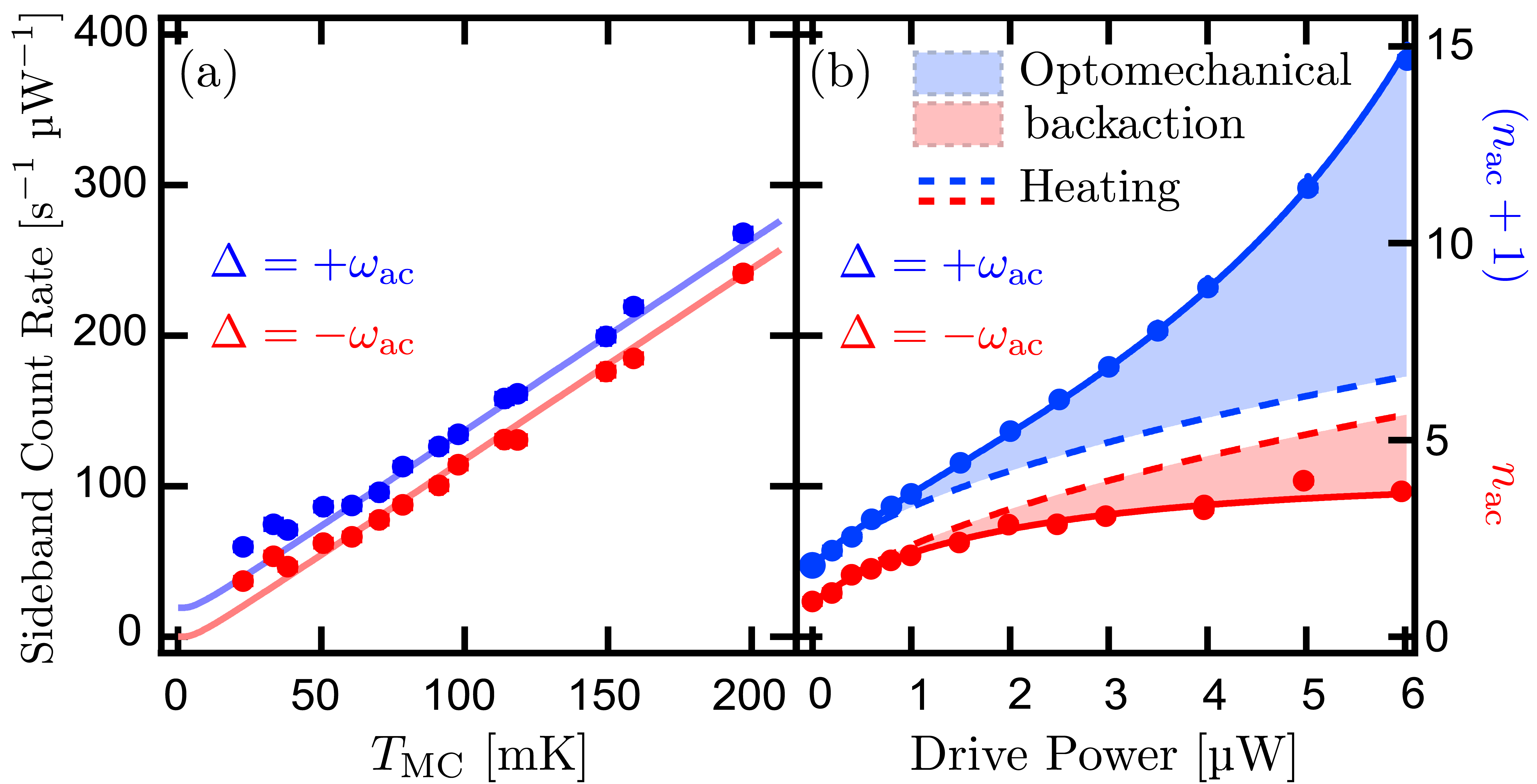}
\caption{ \textbf{Temperature- and power-dependence of the sidebands. } Normalized anti-Stokes (red) and Stokes (blue) photon detection rates. (a) Measurements as a function of $T_\text{MC}$ with $P_\text{in}=250$ nW. Solid lines: fit to the form predicted for thermal equilibrium. (b)  Measurements as a function of $P_\text{in}$ with $T_\text{MC} \approx 20 \text{mK}$. Solid lines: a fit including optomechanical backaction (colored regions) and heating due to optical absorption (dashed lines). Right-hand axis: the mean phonon occupancy $n_\text{ac}$ ($n_\text{ac}+1$) inferred from the detection rates.
}
\label{TempScan}
\end{figure}

Figure \ref{TempScan}(a) shows $R_\text{AS}$ and $R_\text{S}$ measured as a function of $T_\text{MC}$ at an incident drive power $P_\text{in}=250$ nW. The values of  $R_\text{AS}$ and $R_\text{S}$  are extracted from fitting five-point spectra as described in Sec.~\ref{background count}.
Both $R_\text{AS}$ and $R_\text{S}$ increase with $T_\text{MC}$ as expected, and the persistent difference between the two is consistent with the quantum sideband asymmetry (QSA).
The solid lines in Fig.~\ref{TempScan}(a) show a one-parameter fit to the form $a(e^{\hbar\omega_\text{ac}/k_B T_\text{MC}}-1)^{-1}$ (for the red data) and $1+a(e^{\hbar\omega_\text{ac}/k_B T_\text{MC}}-1)^{-1}$ (for the blue data), where  the fit parameter $a=\gamma_\text{ac} C \eta_\kappa\eta_\text{det}/P_{\mathrm{in}}$ is the magnitude of the power-normalized QSA.
The fit only uses data with $T_\text{MC}>50$ mK, as the calibration of the $\text{RuO}_2$ thermometer used to determine $T_\text{MC}$ is uncertain at lower temperatures.

The agreement between data and theory in Fig.~\ref{TempScan}(a) indicates that the mean phonon occupancy $n_\text{ac}$ is primarily determined by $T_\text{MC}$.
However, at higher drive powers, both optical backaction and absorption-induced heating (in the fibers and mirrors) can alter $n_\text{ac}$.
To characterize the role of backaction and heating in these devices, Fig.~\ref{TempScan}b shows $R_\text{AS(S)}$ as a function of $P_\text{in}$.
(The data at $P_\text{in}=0$ are measured using pulsed laser excitations and modelling (and then subtracting) the transient heating.)
For sufficiently low $P_\text{in}$ ($\lesssim$ \SI{300}{\nano\watt}), the values of $R_\text{AS(S)}$ are consistent with $T_\text{MC} \approx 20$ mK, and show a splitting that is dominated by the QSA. 
As $P_\text{in}$ is increased, both $R_\text{AS}$ and $R_\text{S}$ increase.
However, $R_\text{S}$ increases more rapidly with $P_\text{in}$ than $R_\text{AS}$.
This is consistent with the combined effects of heating from optical absorption (which increases $n_\text{ac}$ independent of $\Delta$) and optomechanical backaction (which increases (decreases) $n_\text{ac}$ when $\Delta$ is positive (negative)).

\setlength{\abovecaptionskip}{0pt plus 1pt minus 1pt} 

To analyze these effects quantitatively, we apply the standard theory of optomechanical backaction, and model the heating by assuming that the device is subject to a heatload proportional to $P_{\text{in}}$ and is thermally linked to the MC with a thermal conductance $\sigma = b T^{k+1}$. Before describing this approach in detail (see the following paragraphs), we first summarize the result. The solid lines in Fig.~\ref{TempScan}(b) show the best fit while taking $g_0, T_\text{MC}, \eta_{\text{det}}, b$ and $k$ as fit parameters.
The dashed lines show the change in $n_\text{ac}$ attributable to absorption-induced heating. The best-fit value $k=1.05(3)$ suggests a predominantly metallic thermal conductance between the device and MC ($k=1$ for metals).
The fit also gives a single photon coupling rate $g_0/2\pi=4.58(2)$ kHz and a net detection efficiency of $\eta_{\text{det}}=0.18(1)$.
The best-fit value $T_{\text{MC}}=19(1)$ mK corresponds to a mean thermal phonon occupancy of $n_\text{ac}=0.83(2)$.

In more detail, we start with the power-normalized sideband count rates as described by standard quantum optomechanics theory:
\begin{align}
    \overline{R}_\text{AS} &= a n_\text{ac} \label{EqPowerFitRed} \\
    \overline{R}_\text{S} &= a (n_\text{ac}+1) \label{EqPowerFitBlue}
\end{align}
where, as defined above, $a = \gamma_\text{ac} C \eta_\kappa \eta_\text{det} / P_\text{in}$, and $n_\text{ac}$ is determined by the joint effects from the thermal bath ($n_\text{th}$) and optomechanical backaction \cite{marquardtPRL2007},
\begin{align}
    n_\text{ac} = \frac{ 
    \gamma_\text{opt} \left(\frac{\kappa_\text{c}}{4\omega_\text{ac}}\right)^{2} + \gamma_\text{ac}n_\text{th}}
    {\gamma_\text{opt}+\gamma_\text{ac}}
\label{EqNFinal}
\end{align}
where $\gamma_\text{opt} \left(= \gamma_\text{ac}C = 4 \cdot g_0^2/\kappa_\text{c} \cdot n_\text{c}\right)$ and $\gamma_\text{ac}$ are the optomechanical damping rate and bare mechanical damping rate, respectively, and $n_\text{th}$ accounts for the effect of optical-absorption-induced heating in the fibers and mirrors.

To model the effect of optical heating, we assume a heat load proportional to $P_\text{in}$, and that the conductivity of the thermal link between the fiber and the MC is a power law in temperature with exponent $k$, such that the fiber's temperature is given by
\begin{align}
   T_\text{fib} = (T_\text{MC}^{k+1} + \beta^{k+1}P_\text{in})^{1/(k+1)} 
\label{EqTFib}
\end{align}
where $\beta$ characterizes the amount of heat generated by a given $P_\text{in}$. 
In addition, we assume a ballistic transport of heat (phonons) from the Helium inside the OMC to the MC bath via the Helium channel connecting them.
It follows that
\begin{align}
    n_\text{th}=\frac{n_\text{fib}\gamma_\text{ac0} + n_\text{MC}\gamma_\text{ball}}{\gamma_\text{ac0}+\gamma_\text{ball}}
\label{EqNThermal}
\end{align}
where $n_\text{fib}$ = 1/($e^{{\hbar\omega_\text{ac}}/\left({k_\text{B}T_\text{fib}}\right)}-1)$ and $n_\text{MC}$ = 1/($e^{{\hbar\omega_\text{ac}}/\left({k_\text{B}T_\text{MC}}\right)}-1)$. The bare mechanical damping rate $\gamma_\text{ac}$ = $\gamma_\text{ac0}$ + $\gamma_\text{ball}$ includes: i) the intrinsic mechanical damping rate $\gamma_\text{ac0}$/2$\pi$ = 3.2$\times$10$^{3}$ Hz at zero temperature due to acoustic loss into the fibers, and ii) the three phonon loss $\gamma_\text{ball}$/2$\pi$ = 2.7$\times$10$^{6}$ $T_\text{MC}^{4}$ Hz/K$^4$ at the base MC temperature. 

\setlength{\abovecaptionskip}{0pt plus 2pt minus 2pt} 
\begin{figure}
\centering
\includegraphics[width=0.48\textwidth]{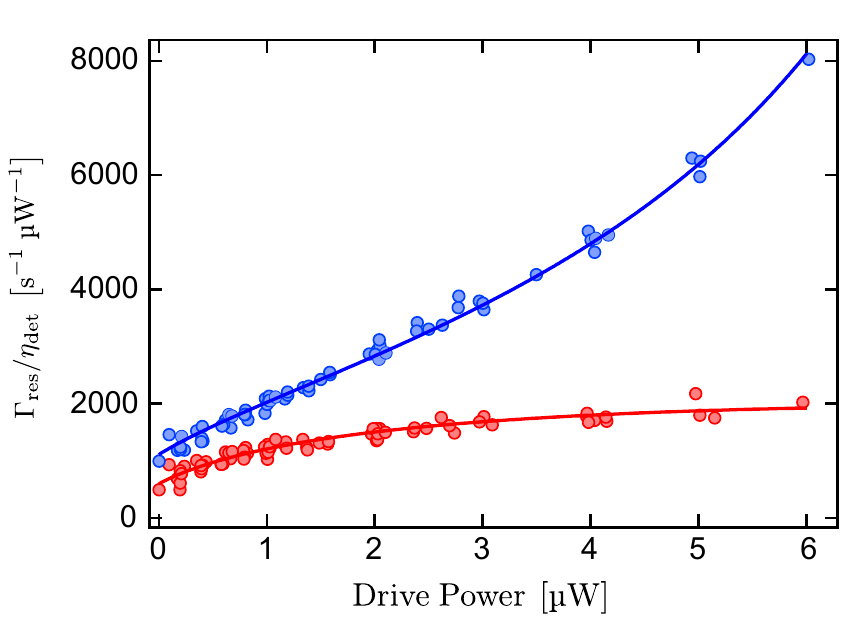}
\caption{
\textbf{Several scans of the drive power.} Scattering rates as a function of drive power, taken over several days. The points are data, and the solid lines are a fit in which all the fit parameters are global, but the detection efficiency is allowed to vary between the data sets. For this plot, each data set is normalized by its fitted detection efficiency.
}
\label{FigPowerFit}
\end{figure}

We have verified the accuracy of this model by using it to analyze drive power sweep data taken on several different days over the course of ten months. Fig.~\ref{FigPowerFit} shows seven different power sweeps. The solid lines are the best fit of the data to the model described by Eq.~\ref{EqPowerFitBlue} - \ref{EqNThermal}. The fit uses a single $T_\text{MC}$, $\beta$, $k$, and optomechanical coupling $g_\text{0}$ for all of the sweeps but allows $\eta_\text{det}$ to vary between the sweeps, as the overall detection efficiency is observed to differ from day to day. 

 The data and the fit are normalized to an ideal detection efficiency (i.e., $\eta_\text{det}$ = 1) for display in Fig.~\ref{FigPowerFit}. The fit gives an initial device temperature of 24.4 $\pm$ 0.4 mK, equivalent to 1.61 $\pm$ 0.03 phonons. The thermal conductivity follows a power law in temperature with an exponent $k = 1.09 \pm 0.03$. The fitting parameters are summarized in Table.~\ref{TableTempModel}.

\setlength{\abovecaptionskip}{0pt plus 2pt minus 2pt} 
\begin{table}[]
    \centering
    \vspace{15pt}
    \begin{tabular}{|c|l|}
        \hline
        $T_\text{MC}$ &~ 24.4 $\pm$ 0.4 ~mK \\
        \hline
        $\beta$ &~  0.54 $\pm$ 0.03 K/$\text{W}^{1/(k+1)}$ \\
        \hline
        $k$ &~  1.09 $\pm$ 0.03\\
        \hline
    \end{tabular}
    \vspace{10pt}
    \caption{
    \textbf{Fit parameters for thermal transport from fiber to bath} The best-fit in Fig.~\ref{FigPowerFit} returns the fit parameters that describe the fiber temperature according to Eq.~\ref{EqTFib}. $T_{\mathrm{MC}}$ characterizes the initial device temperature, $k$ the exponent of the power law that describes the thermal conductivity, and $\beta$ the fraction of heat generated by a given $P_{in}$.
    }
    \label{TableTempModel}
\end{table}

\clearpage

\subsection{Filtering of Counts Detected by the SNSPDs} \label{filte_afterpusling}

\begin{figure}
    \centering
    \includegraphics[scale=0.92]{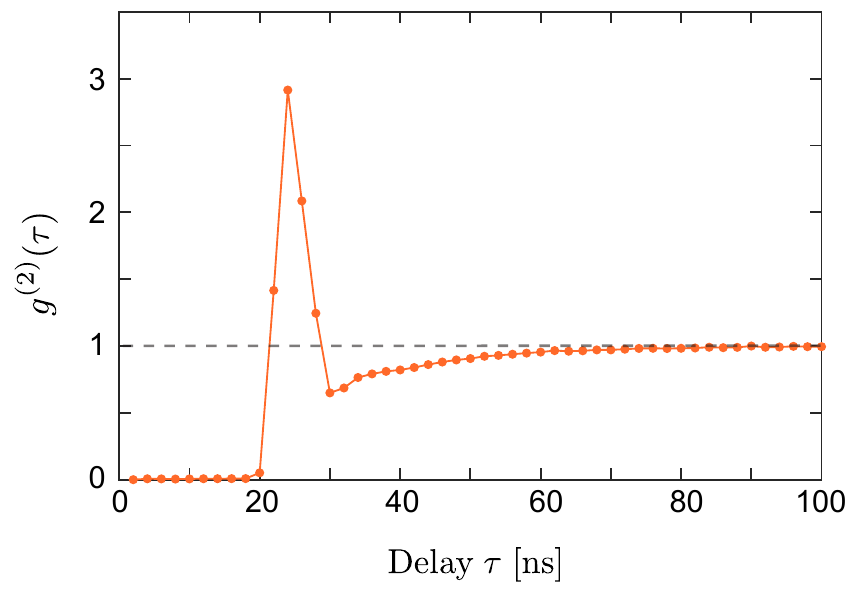}
    \caption{\textbf{Afterpulses in the SNSPDs.} The second-order correlation function of a coherent laser shows a sharp feature in the first 50 ns due to the afterpulsing effect. The flat region before 20 ns is the dead time of the SNSPDs. For delays $> 30$ ns,  the correlation decays to 1, as expected for a coherent state.}
    \label{fig:afterpulse}
\end{figure}

Not every count registered by the SNSPDs corresponds to the detection of a photon. 
Here we cosider three types of false-counts: (i) afterpulses, (ii) rapid bursts of counts in a short duration (10 - \SI{100}{\micro\second}), and (iii) SNSPD dark counts. 
This section describes the filtering protocols used to prevent false-counts of types (i) and (ii) from compromising the coherence measurements.
False counts of type (iii) (i.e., dark counts) cannot be identified or filtered, given that they occur at random times and are uniformly distributed in time; however, their effect on measured coherences can be accounted for, as is detailed in Sec.~\ref{background photon}.

\textbf{(i) Afterpulses} It is well-known that reflection of the voltage pulses generated by SNSPDs can cause false counts \cite{burenkov2013investigations,fujiwara2011afterpulse}.
Following a detection event at time $t=0$, an afterpulse can occur at a delay $\tau_\text{ap}$ set by the time it takes for the voltage reflection to reach the SNSPD.

The signature of afterpulses is illustrated in Fig.~\ref{fig:afterpulse}. It shows the second-order coherence $g^{(2)}(\tau)$ of a power-stabilized laser, measured with a bin-size of $2$ ns. The peak at delay $\tau_\text{ap}\approx 24$ ns is caused by the afterpulses. For $\tau>50$ ns, $g^{(2)}(\tau) \approx 1$, as expected for laser light. 

To reduce the impact of these afterpulses in the experiments described here, any count that is registered within $50$ ns of a preceding count (on a given SNSPD) is tagged as an afterpulse and discarded.
This also removes any true counts that occur within $50$ ns of each other; however the error this introduces in coherence measurements is to effectively shorten the bin at the smallest delay by $50$ ns. This can easily be compensated for, but since the shortest delays used in this work are $\tau\ge$ \SI{1}{\micro\second}, we need not perform such a compensation. For the same reason, we also need not compensate for the recovery time of the SNSPD, which is specified to be $50$ ns.

\textbf{(ii) Rapid Bursts} There occur rare instances ($<0.008\%$ of all 90 ms DAq configuration intervals, c.f. Sec.~\ref{opticalSetupSection}) in which the SNSPDs register a periodic train of counts with an abnormally high rate ($\gtrsim 10^3\times$ expected count rate) for a short duration ($10-100$ \textmu s).
While the underlying cause of these spurious counts is unknown, such events can be identified as outliers and discarded. This is done by using a statistical model which tags events having an extremely low probability of occurring, as described below.

Suppose the mean count rate during an experiment is measured to be $I$. The average number of counts received in a time interval $\Delta t$ is thus $\lambda=I\Delta t$. The probability of receiving $k$ photons in this interval can be estimated using a statistical model as $P(k,\lambda)$. Thus, for a total DAq time $T$, among the $N=T/\Delta t$ number of $\Delta t$- intervals, the mean number of intervals in which $k$ counts are recorded is expected to be $N \times P(k,\lambda)$. To reject outliers, we set a threshold $\epsilon \ll 1$ and search for all $\Delta t$-intervals that receive $\geq k_\text{thr}$ counts, where $k_\text{thr}$ is the smallest value of $k$ for which $N \times P(k_\text{thr},\lambda) < \epsilon$. For each $\Delta t$-interval that is identified as an outlier in this way, we discard the entire $90$ ms DAq record containing this interval. In this work, $N\sim 10^8-10^{13}$ and $\epsilon$ is set to $0.1$.

\setlength{\abovecaptionskip}{0pt plus 2pt minus 2pt}
\begin{figure}[]
    \centering
    \includegraphics[scale=0.92]{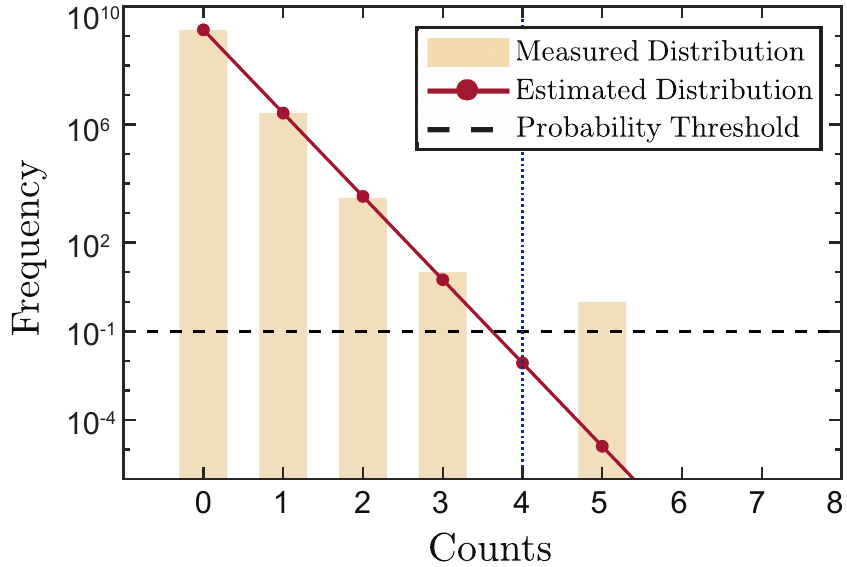}
    \caption{\textbf{Distribution of counts.} Number of intervals as a function of counts. We binned data into 3 $\mathrm{\mu s}$ intervals. The beige bars are the histogram of counts in the intervals. The red solid line is the estimated distribution of counts in the short interval limit $\Delta t\ll\tau_\text{c}$, where $\tau_{\mathrm{c}}$ is the coherence time. We used a probability threshold $0.1$ (shown as the black dashed line)  corresponding to a count threshold $k_{\mathrm{thr}}$ (shown by the blue dotted line). In this data, one interval with 5 counts is rejected by this protocol.}
    \label{count stat}
\end{figure}

Figure \ref{count stat} illustrates this protocol.
It shows the experimentally measured frequency with which $k-$counts were detected in $\Delta t=$ \SI{3}{\micro\second} intervals for $T\approx 4800$ s ($N\sim 1.6\times 10^9$) and $I\approx 2500$ counts/s ($\lambda\sim 1.51\times 10^{-3})$.
The red solid line represents the statistical model $P(k,\lambda)$ (see below).
For $k\ge 4$, $N\times P(k,\lambda)<0.1$, so that $k_\text{thr}=4$.
As such, the $90$ ms DAq intervals that included any $\Delta t=$ \SI{3}{\micro\second} intervals that registered $\ge 4$ counts were discarded from the dataset.
In this instance, $1$ out of $53,421$ DAq intervals was discarded.

We iterated the filtering protocol illustrated above for $\Delta t= 3, 10, 30, 100$ and \SI{300}{\micro\second}.
The best apriori statistical guess for $P(k,\lambda)$ depends on the choice of $\Delta t$, the measured count rate $I$, and the coherence timescale $\tau_\text{c}$ of the photon source.
For $\Delta t\gg \tau_\text{c}$ and $I^{-1}\ll \tau_\text{c}$, the photons detected within the interval $\Delta t$ can be approximated as uncorrelated, and we model $P(k,\lambda)$ as a Poissonian, $\text{Pois}(k,\lambda)=\lambda^k e^{-\lambda}/k!$ \cite{kelley1964theory}.
However, if $\Delta t\lesssim \tau_\text{c}$ or $I^{-1} \gtrsim \tau_\text{c}$, $P(k,\lambda)$ is best modelled by accounting for the coherence of the source.
Given that the acoustic mode generating the sideband photons is expected to be in thermal equilibrium, $P(k,\lambda)$ is well modelled by $P_\text{th}(k,\lambda) = \lambda^k / (1+\lambda)^{k+1}$ for $\Delta t \ll \tau_\text{c}$.
For $\Delta t \sim \tau_\text{c}$, $P(k,\lambda)$ can be estimated to be between these two limiting cases of $\Delta t\gg \tau_\text{c}$ and $\Delta t\ll \tau_\text{c}$.
To avoid making an apriori guess for $\tau_\text{c}$, we chose to model $P(k,\lambda)$ with $P_\text{th}(k,\lambda)$ for all $\Delta t$. This achieves a more conservative filtering of the data as it allows for more frequent occurrences of a higher number of counts within any $\Delta t$-interval ($P_\text{th}(k,\lambda) > P(k,\lambda)$ for $k>2, \lambda < 1$).

Applying this protocol to all of the data acquired used in this work, $133$ out of $\sim1.86\times 10^6$ DAq intervals were tagged and discarded as outliers.

\subsection{Subtraction of Coincidences due to Background Photons}\label{background photon}
This section describes the method used to correct the correlation functions for the presence of background counts in the SNSPD data. As described in Sec.~\ref{background count},
these background counts result from stray light, leakage through the filter cavities, and the SNSPD's dark counts.

We measure (not shown) these background counts to have a fixed mean arrival rate over the duration of the experiments, and to be uncorrelated among themselves, i.e. they are Poisson distributed.
We also assume them to be independent of the sideband photons.
Therefore, the measured second-order correlation function $g^{(2)}_{\mathrm{exp}}(\tau)$, including the cross-correlations between the sideband photons and the background photons, is given by
\begin{align*}
    g^{(2)}_\text{exp}(\tau) 
    &=\frac{\langle 
        [a^\dagger + \xi^\dagger](0) 
        [a^\dagger + \xi^\dagger](\tau) 
        [a + \xi](\tau)
        [a + \xi](0) 
        \rangle}
        {\langle [a^\dagger + \xi^\dagger](0) [a + \xi](0) \rangle
         \langle [a^\dagger + \xi^\dagger](\tau) [a + \xi](\tau) \rangle }\\
    &=\frac{g^{(2)}(\tau) + 2\epsilon + \epsilon^2}{ (1 + \epsilon)^2 },
\end{align*}
where $a$ and $\xi$ correspond to the sideband and background photons, respectively, and we have used $ \langle a^\dagger\xi \rangle = 0 = \langle \xi^\dagger a\rangle$, and $\langle \xi^\dagger(0) \xi^\dagger(\tau) \xi(\tau) \xi(0) \rangle = \langle \xi^\dagger \xi \rangle^2$.
$\epsilon = \langle \xi^\dagger \xi \rangle / \langle a^\dagger a \rangle $ is the ratio of the mean background and sideband count rates.
For the drive powers used in this work, $0.04\lesssim\epsilon\lesssim0.2$.

The corrected second-order correlation function $g^{(2)}(\tau)$ is thus given by
\begin{align*}
    g^{(2)}(\tau)=g^{(2)}_{\mathrm{exp}}(\tau)+2(g^{(2)}_{\mathrm{exp}}(\tau)-1)\epsilon+(g^{(2)}_{\mathrm{exp}}(\tau)-1)\epsilon^2.
\end{align*}
A similar calculation gives the corrected third-order and fourth-order correlation functions using
\begin{widetext}
\begin{align*}\label{third-order correlation correction}
    g^{(3) }_{\mathrm{exp}}(\tau_1,\tau_2) = 
    \dfrac{
    g^{(3)}(\tau_1,\tau_2)+\epsilon\left(g^{(2)}(\tau_1)+g^{(2)}(\tau_2)+g^{(2)}(\tau_1+\tau_2)\right)+3\epsilon^2+\epsilon^3}
    {(1+\epsilon)^3},\\
    g^{(4) }_{\mathrm{exp}}(\tau_1,\tau_2,\tau_3) = 
    \dfrac{g^{(4)}(\tau_1,\tau_2,\tau_3)+\epsilon\sum_{g_{i}^{(3)} \in \mathbb{G}^{(3)}} g_{i}^{(3)}+\epsilon^2\sum_{g_{j}^{(2)} \in \mathbb{G}^{(2)}} g_{j}^{(2)}+4\epsilon^3+\epsilon^4}
    {(1+\epsilon)^4},
\end{align*}
where $\mathbb{G}^{(3)}$ and $\mathbb{G}^{(2)}$ are
\begin{align*}
\mathbb{G}^{(3)}&=\set{g^{(3)}(\tau_1,\tau_2),g^{(3)}(\tau_1+\tau_2,\tau_3),g^{(3)}(\tau_1,\tau_2+\tau_3),g^{(3)}(\tau_2,\tau_3)},\nonumber\\
\mathbb{G}^{(2)}&=\set{g^{(2)}(\tau_1),g^{(2)}(\tau_2),g^{(2)}(\tau_3),g^{(2)}(\tau_1+\tau_2),g^{(2)}(\tau_2+\tau_3),g^{(2)}(\tau_1+\tau_2+\tau_3)}.\nonumber
\end{align*}
\end{widetext}
~

\clearpage

\subsection{Correspondence between Photon and Phonon Correlation Functions}  \label{photonphonon}
In the main text, we introduced the normally ordered phonon coherence $ g_\text{ac}^{(n)}$ obtained from the anti-Stokes photon coherence and the anti-normally ordered phonon coherence $ h_\text{ac}^{(n)}$ obtained from the Stokes photon coherence. In this section, we formally demonstrate this one-to-one correspondence between coherences of the optical fields output by the cavity and the coherences of the phonons. We also note that optical losses along the detection chain do not compromise these coherence measurements \cite{riedinger2016}.

In a frame rotating at the optical drive frequency $\omega_\text{D} = \omega_\text{c} + \Delta$, the output optical field $o$ of the optomechanical cavity has a spectrum $S_{o^\dagger o}$ that relates to mechanical spectrum $S_{xx}$ as \cite{borkje2011}
\begin{equation}
    S_{o^{\dagger} o} (\omega)
    = \kappa_\text{in} n_\text{c} 2\pi \delta(\omega) + \kappa_\text{in} g_0^2 n_\text{c} |\chi_\text{c}(-\omega)|^2 S_{xx}(\omega), \label{EqSoo}
\end{equation}
where the first term corresponds to the optical drive field and the second term corresponds to the optical fields produced by the mechanical motion through the optomechanical coupling.
$\chi_\text{c}(\omega)$ is the bare optical cavity susceptibility $\chi_\text{c}(\omega)= (\kappa_\text{c}/2 -i(\omega+\Delta))^{-1}$.
The field $p$ detected by the SNSPDs is the cavity output field $o$ passed through the filter cavities with transmission $f_\text{filter}(\omega)$,
\begin{equation*}
    p(t) = \int\limits_{-\infty}^\infty \frac{d\omega}{2\pi} e^{-i\omega t}f_\text{filter}(\omega) o(\omega),
\end{equation*}
so that the detected normally ordered optical correlations are given by the Weiner-Khinchin relation
\begin{align*}
    \langle p^\dagger(t+\tau) p(t) \rangle = \int\limits_{-\infty}^\infty \frac{d\omega}{2\pi}~ e^{-i\omega\tau} |f_\text{filter}(-\omega)|^2 S_{o^\dagger o}(\omega),
\end{align*}
$f_\text{filter}(\omega)$ is non-zero only around $\omega = +\omega_\text{ac}$ for the red-detuned drive (and only around $\omega = -\omega_\text{ac}$ for the blue-detuned drive), within the two filter cavity linewidths $\kappa_\text{FC1}, \kappa_\text{FC2} \ll \omega_\text{ac}$.
This ensures that the contribution of the first term in Eq.~\ref{EqSoo} to this integral, corresponding to the contribution of the drive field, is negligible. As such, 
\begin{align}
    &\langle p^\dagger(t+\tau) p(t) \rangle = \label{EqPdaggerP} \\
    &\kappa_\text{in} g_0^2 n_\text{c} \int\limits_{-\infty}^\infty \frac{d\omega}{2\pi}~ e^{-i\omega\tau} |f_\text{filter}(-\omega)|^2 |\chi_\text{c}(-\omega)|^2 S_{xx}(\omega). \nonumber
\end{align}
The contribution of $S_{xx}(\omega)$ to these optical correlations can be decomposed as 
\begin{align*}
    S_{xx}(\omega) = x_\text{zpt}^2 [S_{b^\dagger b}(\omega) + S_{b b^\dagger}(\omega)],
\end{align*}
with $x = x_\text{zpt} (b+b^\dagger)$, and $x_\text{zpt}$ being the zero-point motion of the oscillator.
$S_{b^\dagger b}(\omega)$ is non-zero only around $\omega = -\omega_\text{ac}$, and $S_{b b^\dagger}(\omega)$ is non-zero only around $\omega = +\omega_\text{ac}$, corresponding to the anti-Stokes and Stokes scattering processes respectively, which occur within the mechanical susceptibility $\chi_\text{ac}(\omega)$ of linewidth $\bar\gamma_\text{ac}$.
This is explicitly seen through the oscillator's Langevin equation \cite{lemondePRA2014, GardinerPRA1985, WiessQuantumDissipativeSystems, loudonBook}
\begin{align*}
    \dot b &= - \left(\frac{\bar\gamma_\text{ac}}{2} -i\omega_\text{ac} \right) b + \sqrt{\bar\gamma_\text{ac}} b_\text{in},
\end{align*}
where $b_\text{in}$ is the thermal noise input with $\langle b_\text{in} \rangle = 0$.
The correlations for $\tau \ge 0$ are given by $\langle b_\text{in}^\dagger(\tau) b_\text{in}(0) \rangle = n_\text{ac} \delta(\tau)$ and $\langle b_\text{in}(\tau) b_\text{in}^\dagger(0) \rangle = (n_\text{ac}+1) \delta(\tau)$.
And $n_\text{ac}=\langle b(t)^\dagger b(t)\rangle$ is assumed to be time-independent (stationary). The formal integral
\begin{equation*}
    b(\tau) = \sqrt{\bar\gamma_\text{ac}} \int\limits_{-\infty}^\infty dt' 
    e^{- (\bar\gamma_\text{ac}/2-i\omega_\text{ac}) (\tau-t')} b_\text{in}(t')
\end{equation*}
yields the correlations and spectra as
\begin{equation}
    \langle b^\dagger(\tau) b(0) \rangle = n_\text{ac} e^{-(\bar\gamma_\text{ac}/2+i\omega_m)\tau}, \label{EqCorr1}
\end{equation}
\begin{align*}
    S_{b^\dagger b}(\omega) &= \int\limits_{-\infty}^\infty d\tau~ e^{i\omega\tau} \langle b^\dagger(\tau) b(0) \rangle\\
    &= n_\text{ac} \bar\gamma_\text{ac} |\chi_\text{ac}(-\omega)|^2\\
    &= \frac{ n_\text{ac} \bar\gamma_\text{ac} } {(\bar\gamma_\text{ac}/2)^2+(-\omega-\omega_\text{ac})^2}.
\end{align*}
And similarly
\begin{equation}
    \langle b(\tau) b^\dagger(0) \rangle = (n_\text{ac} + 1) e^{-(\bar\gamma_\text{ac}/2-i\omega_m)\tau}, \label{EqCorr2}
\end{equation}
\begin{align*}
    S_{b b^\dagger}(\omega) &= \int\limits_{-\infty}^\infty d\tau~ e^{i\omega\tau} \langle b(\tau) b^\dagger(0) \rangle\\
    &= (n_\text{ac}+1) \bar\gamma_\text{ac} |\chi_\text{ac}(\omega)|^2\\
    &= \frac{ (n_\text{ac}+1) \bar\gamma_\text{ac} } {(\bar\gamma_\text{ac}/2)^2+(\omega-\omega_\text{ac})^2}.
\end{align*}

Thus, for the case of the red-detuned drive $\Delta = -\omega_\text{ac}$, only the $S_{b^\dagger b}$ term contributes to the integral around $\omega = -\omega_\text{ac}$ in Eq.~\ref{EqPdaggerP}, corresponding to the anti-Stokes scattered photons, and yields
\begin{align*}
    &\langle p^\dagger(t+\tau) p(t) \rangle = \langle a_\text{AS}^\dagger(t+\tau) a_\text{AS}(t) \rangle \\
    &= \kappa_\text{in} g_0^2 n_\text{c} |\chi_\text{c}(\omega_\text{ac})|^2 x_\text{zpt}^2 \int\limits_{-\infty}^\infty \frac{d\omega}{2\pi}~ e^{-i\omega\tau} S_{b^\dagger b}(\omega)\\
    &=4\kappa_\text{in} g_0^2 n_\text{c} x_\text{zpt}^2 / \kappa_\text{c}^2 ~ \langle b^\dagger(t+\tau) b(t) \rangle,
\end{align*}
where we have approximated $f_\text{filter}(-\omega) \approx 1$ and $\chi_\text{c}(-\omega) \approx \chi_\text{c}(\omega_\text{ac})$ for $\omega$ around $-\omega_\text{ac}$, as $\bar\gamma_\text{ac} \ll \kappa_\text{FC1}, \kappa_\text{FC2} \ll \kappa_\text{c}$.

For the case of the blue-detuned drive $\Delta = +\omega_\text{ac}$, only the $S_{b b^\dagger}$ term contributes to the integral around $\omega = +\omega_\text{ac}$ in Eq.~\ref{EqPdaggerP}, corresponding to the Stokes scattered photons, and yields
\begin{align*}
    &\langle p^\dagger(t+\tau) p(t) \rangle = \langle a_\text{S}^\dagger(t+\tau) a_\text{S}(t) \rangle \\
    &= \kappa_\text{in} g_0^2 n_\text{c} |\chi_\text{c}(-\omega_\text{ac})|^2 x_\text{zpt}^2 \int\limits_{-\infty}^\infty \frac{d\omega}{2\pi}~ e^{-i\omega\tau} S_{b b^\dagger}(\omega)\\
    &=4\kappa_\text{in} g_0^2 n_\text{c} x_\text{zpt}^2 / \kappa_\text{c}^2 ~ \langle b(t+\tau) b^\dagger(t) \rangle,
\end{align*}
where we have approximated $f_\text{filter}(-\omega) \approx 1$ and $\chi_\text{c}(-\omega) \approx \chi_\text{c}(-\omega_\text{ac})$ for $\omega$ around $+\omega_\text{ac}$, as $\bar\gamma_\text{ac} \ll \kappa_\text{FC1}, \kappa_\text{FC2} \ll \kappa_\text{c}$.

This demonstrates the correspondence between the detected optical correlations and the appropriately ordered phonon correlations.
Note that high-order correlations can be evaluated in terms of such two-time correlations using Wick's theorem, for any Gaussian state, thereby implying the correspondence of the detected optical coherences and the appropriately ordered phonon coherences.
Note that the proportionality constant $(4 \kappa_\text{in} g_0^2 n_\text{c} x_\text{zpt}^2 / \kappa_\text{c}^2)$ is immaterial for the coherences, which are normalized ratios of such correlations.

\subsection{Correlation Functions of a Thermal State}\label{TheoryCorrelations}
This section presents calculations of coherence functions for a thermal state of an oscillator. These functions are used to fit the data in the main text.

We calculate the coherences for an oscillator's thermal state using Wick's theorem. 
For any Gaussian state, the theorem implies that high-order moments can be evaluated in terms of second moments.
For example, the numerators in 
\begin{equation*}
    g^{(2)}_\text{ac}(\tau_1) = \frac{\langle b_0^\dagger b_1^\dagger b_1 b_0 \rangle}{\langle b_0^\dagger b_0 \rangle \langle b_1^\dagger b_1 \rangle},~
    h^{(2)}_\text{ac}(\tau_1) = \frac{\langle b_0 b_1 b_1^\dagger b_0^\dagger \rangle}{\langle b_0 b_0^\dagger \rangle \langle b_1 b_1^\dagger \rangle}
\end{equation*}
where $b_0 = b(0)$ and $b_1 = b(\tau_1)$, evaluate as
\begin{align*}
    \langle b_0^\dagger b_1^\dagger b_1 b_0 \rangle &= 
    \langle b_0^\dagger b_0 \rangle \langle b_1^\dagger b_1 \rangle +
    \langle b_0^\dagger b_1 \rangle \langle b_1^\dagger b_0 \rangle,\\
    \langle b_0 b_1 b_1^\dagger b_0^\dagger \rangle &= 
    \langle b_0 b_0^\dagger \rangle \langle b_1 b_1^\dagger \rangle +
    \langle b_0 b_1^\dagger \rangle \langle b_1 b_0^\dagger \rangle.
\end{align*}

\setlength{\abovecaptionskip}{0pt plus 2pt minus 2pt}
\begin{figure*}
\centering
\includegraphics[width=0.98\textwidth]{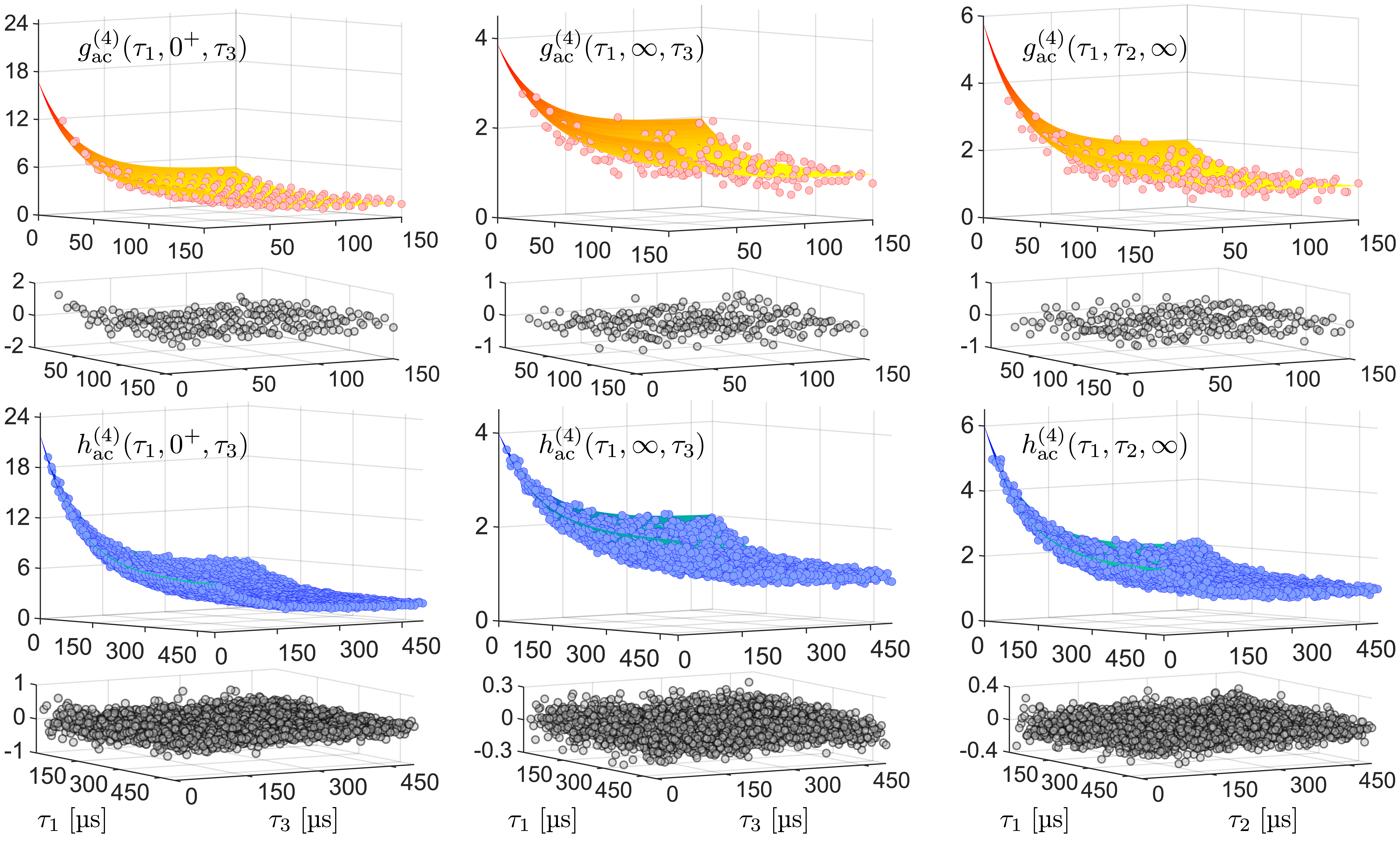}
\caption{\textbf{
Fourth-order phonon coherences.} $g^{(4)}_\text{ac}(\tau_1,\tau_2,\tau_3)$ and $h^{(4)}_\text{ac}(\tau_1,\tau_2,\tau_3)$ measured for $P_\text{in}\approx$ \SI{5}{\micro\watt}, with photon arrival times binned in \SI{10}{\micro\second} bins. Shown are three sets of representative 2D slices: $g^{(4)}_\text{ac}(\tau_1,0^+,\tau_3)$ and $h^{(4)}_\text{ac}(\tau_1,0^+,\tau_3)$, $g^{(4)}_\text{ac}(\tau_1,\infty,\tau_3)$ and $h^{(4)}_\text{ac}(\tau_1,\infty,\tau_3)$, $g^{(4)}_\text{ac}(\tau_1,\tau_2,\infty)$ and $h^{(4)}_\text{ac}(\tau_1,\tau_2,\infty)$, where $\tau=0^+$ represents the bin with $\SI{5}{\micro\second}<\tau<\SI{15}{\micro\second}$ and $\tau=\infty$ represents times $\tau>10 \bar\gamma_\text{ac}^{-1}$. Fits are to the entire 3D ($\tau_1$-, $\tau_2$-, $\tau_3$- dependent) data. Fit residuals are shown in black.
}
\label{Fig10}
\end{figure*}

For the thermal state, which is a Gaussian state, these second moments are given for $\tau \ge 0$ by Eq.~\ref{EqCorr1}, \ref{EqCorr2}, i.e. 
\begin{align*}
    \langle b^\dagger(\tau) b(0) \rangle &= n_\text{ac} e^{-(\bar\gamma_\text{ac}/2 + i\omega_\text{ac})\tau},\\
    \langle b(\tau) b^\dagger(0) \rangle &= (n_\text{ac}+1) e^{-(\bar\gamma_\text{ac}/2 - i\omega_\text{ac})\tau}.
\end{align*}
The second moments for $\tau \le 0$ are obtained by conjugation.
Combined, these readily give
\begin{align*}
    g^{(2)}_\text{ac}(\tau) &= 1 + e^{-\bar\gamma_\text{ac}\tau} =  h^{(2)}_\text{ac}(\tau).
\end{align*}
To illustrate further, the numerator in 
\begin{equation*}
    g^{(3)}_\text{ac}(\tau_1,\tau_2) = 
    \frac{\langle b_0^\dagger b_1^\dagger b_2^\dagger b_2 b_1 b_0 \rangle}{\langle b_0^\dagger b_0 \rangle \langle b_1^\dagger b_1 \rangle \langle b_2^\dagger b_2 \rangle},
\end{equation*}
where $b_2 = b(\tau_1+\tau_2)$, evaluates as
\begin{align*}
    &\langle b_0^\dagger b_1^\dagger b_2^\dagger b_2 b_1 b_0 \rangle =
    \langle b_0^\dagger b_0 \rangle \langle b_1^\dagger b_1 \rangle \langle b_2^\dagger b_2 \rangle \\
    &+\langle b_0^\dagger b_0 \rangle \langle b_1^\dagger b_2 \rangle \langle b_2^\dagger b_1 \rangle
    + \langle b_0^\dagger b_1 \rangle \langle b_1^\dagger b_0 \rangle \langle b_2^\dagger b_2 \rangle \\
    &+\langle b_0^\dagger b_1 \rangle \langle b_1^\dagger b_2 \rangle \langle b_2^\dagger b_0 \rangle 
    +\langle b_0^\dagger b_2 \rangle \langle b_1^\dagger b_0 \rangle \langle b_2^\dagger b_1 \rangle\\
    &+\langle b_0^\dagger b_2 \rangle \langle b_1^\dagger b_1 \rangle \langle b_2^\dagger b_0 \rangle.
\end{align*}
This, and evaluations of $h^{(3)}_\text{ac}, g^{(4)}_\text{ac}$, and $ h^{(4)}_\text{ac}$ in a similar manner, gives
\begin{align}
    &g^{(3)}_\text{ac}(\tau_1,\tau_2) = h^{(3)}_\text{ac}(\tau_1,\tau_2)\nonumber\\ 
    &=1+e^{-\bar\gamma_\text{ac}\tau_1}
    +e^{-\bar\gamma_\text{ac}\tau_2}
    +3e^{-\bar\gamma_\text{ac}(\tau_1+\tau_2)}, \nonumber\\
    &g_\text{ac}^{(4)}(\tau_1,\tau_2,\tau_3)= h_\text{ac}^{(4)}(\tau_1,\tau_2,\tau_3) \nonumber\\
    &=1+e^{-\bar\gamma_\text{ac}\tau_1}
    +e^{-\bar\gamma_\text{ac}\tau_2}
    +e^{-\bar\gamma_\text{ac}\tau_3} \nonumber\\
    &+3e^{-\bar\gamma_\text{ac}(\tau_1+\tau_2)}
    +3e^{-\bar\gamma_\text{ac}(\tau_2+\tau_3)}
    +e^{-\bar\gamma_\text{ac}(\tau_1+\tau_3)} \nonumber \\
    &+9e^{-\bar\gamma_\text{ac}(\tau_1+\tau_2+\tau_3)}
    +4e^{-\bar\gamma_\text{ac}(\tau_1+2\tau_2+\tau_3)}. \label{Eq:g4}
\end{align}
Limiting cases amenable to an intuitive interpretation are discussed below.

The fourth-order phonon coherence functions $g^{(4)}_\text{ac}(\tau_1,\tau_2,\tau_3)$ and $h^{(4)}_\text{ac}(\tau_1,\tau_2,\tau_3)$ are three-dimensional data sets that can be visualized as two-dimensional slices. 
In Fig.~2(c) of the main text we show one set of 2D slices $g^{(4)}_\text{ac}(0^+,\tau_2,\tau_3)$ and $h^{(4)}_\text{ac}(0^+,\tau_2,\tau_3)$, where $\tau=0^+$ represents the bin with $\SI{5}{\micro\second}<\tau<\SI{15}{\micro\second}$. 
In Fig.~\ref{Fig10}, we show three more sets of 2D slices:
$g^{(4)}_\text{ac}(\boldsymbol{\tau})$ and $h^{(4)}_\text{ac}(\boldsymbol{\tau})$ measured at 
$\boldsymbol{\tau} = (\tau_1, 0^+,\tau_3)$ (panel (a)),
$\boldsymbol{\tau} = (\tau_1,\infty,\tau_3)$ (panel (b)), and
$\boldsymbol{\tau} = (\tau_1,\tau_2,\infty)$ (panel (c)). 
Here, $\tau=\infty$ represents times $\tau>10 \bar\gamma_\text{ac}^{-1}$. Fits to the entire 3D ($\tau_1$-, $\tau_2$-, $\tau_3$- dependent) data are shown as 2D surfaces, with fit residuals shown in black.

The 2D slice $g^{(4)}_\text{ac}(\tau_1,\infty,\tau_3)$ is equivalent to the product of two one-dimensional $g^{(2)}_\text{ac}(\tau)$ functions. 
If the third photon arrives with delay $\tau_2=\infty$, then the arrivals of the third and fourth photons are uncorrelated with the arrival of the first pair of photons. 
That is, $g^{(4)}_\text{ac}(\tau_1,\infty,\tau_3)$ is proportional to the product of the probability of measuring a pair of photons separated by a delay $\tau_1$ and the probability of measuring another pair of photons separated by a delay $\tau_3$. This can be seen from the expression for $g^{(4)}_\text{ac}(\tau_1,\tau_2,\tau_3)$ given in Eq.~\ref{Eq:g4}, where setting $\tau_2=\infty$ results in 
\begin{align}
g_{\mathrm{ac}}^{(4)}(\tau_1,\infty,\tau_3)&=1+e^{-\bar\gamma_\text{ac}\tau_1}+e^{-\bar\gamma_\text{ac}\tau_3}+e^{-\bar\gamma_\text{ac}(\tau_1+\tau_3)}\nonumber\\
&=g_{\mathrm{ac}}^{(2)}(\tau_1)g_{\mathrm{ac}}^{(2)}(\tau_3)\nonumber.
\end{align}
The same equivalence holds for $h_{\mathrm{ac}}^{(4)}(\tau_1,\infty,\tau_3)=h_{\mathrm{ac}}^{(2)}(\tau_1)h_{\mathrm{ac}}^{(2)}(\tau_3)$.

Similarly, the slice $g^{(4)}_\text{ac}(\tau_1,\tau_2,\infty)$ is equivalent to the two-dimensional $g^{(3)}_\text{ac}(\tau_1,\tau_2)$. 
If the fourth photon arrives with delay $\tau_3=\infty$, then its arrival is uncorrelated with the arrivals of the first three photons. Therefore, $g^{(4)}_\text{ac}(\tau_1,\tau_2,\infty)$ is proportional to the probability of measuring a triplet of photons with a delay $\tau_1$ between the first and second and a delay $\tau_2$ between the second and third. We can also see this by setting $\tau_3=\infty$ in Eq.~\ref{Eq:g4}, which yields
\begin{align}
g_{\mathrm{ac}}^{(4)}(\tau_1,\tau_2,\infty)&=1+e^{-\bar\gamma_\text{ac}\tau_1}+e^{-\bar\gamma_\text{ac}\tau_2}+3e^{-\bar\gamma_\text{ac}(\tau_1+\tau_2)}\nonumber\\
&=g_{\mathrm{ac}}^{(3)}(\tau_1,\tau_2)\nonumber.
\end{align}
Again, the same is true for $h_{\mathrm{ac}}^{(4)}(\tau_1,\tau_2,\infty)=h_{\mathrm{ac}}^{(3)}(\tau_1,\tau_2)$.

\subsection{Coherences of \textit{k}-quanta- Subtracted/Added Thermal States}
In the main text we discussed out-of-equilibrium \textit{k}-quanta-subtracted/added thermal states. In this section, we derive the relation between the mean occupancies and coherences of such states and those of the thermal states they are generated from.

The subtraction (addition) of a single phonon at time $t=0$ from a state described by density matrix $\rho$ yields the state with density matrix $\rho_{-1}$ ($\rho_{+1}$)
\begin{equation}
    \rho_{-1}
    = \frac{b(0)\rho b^\dagger(0)}{\text{Tr}[b(0)\rho b^\dagger(0)]}
    = \frac{b(0)\rho b^\dagger(0)}{\text{Tr}[\rho b^\dagger(0) b(0)]}
    =\frac{b(0)\rho b^\dagger(0)}{\langle b^\dagger(0) b(0) \rangle_{\rho}} \label{EqRho'}
\end{equation}
\begin{equation*}
    \rho_{+1}
    = \frac{b^\dagger(0)\rho b(0)}{\text{Tr}[b^\dagger(0)\rho b(0)]}
    = \frac{b^\dagger(0)\rho b(0)}{\text{Tr}[\rho b(0) b^\dagger(0)]}
    =\frac{b^\dagger(0)\rho b(0)}{\langle b(0) b^\dagger(0) \rangle_{\rho}}
\end{equation*}
where the denominator signifies the normalization, and $\langle \cdot \rangle_{\rho} $ is the expectation value for state $\rho$. 
The coherences and mean phonon number of the single-phonon subtracted (added) state $\rho_{-1}$ ($\rho_{+1}$) can thus be evaluated in terms of the  coherences of the steady state $\rho$, as follows.

\textbf{Coherences} By way of illustration, consider the second-order coherence of the single-phonon subtracted state, 
\begin{equation*}
    g_{\text{ac}}^{(2)}\big|_{-1}(\tau) 
    = \frac{
    \langle b^\dagger(0) b^\dagger(\tau) b(\tau) b(0) \rangle_{\rho_{-1}}
    }{
    \langle b^\dagger(0) b(0) \rangle_{\rho_{-1}} \  
    \langle b^\dagger(\tau) b(\tau) \rangle_{\rho_{-1}}
    }
\end{equation*}
Using Eq.~\ref{EqRho'}, the numerator equates to
\begin{equation*}
    \frac{
    \langle b^\dagger(0) b^\dagger(0) b^\dagger(\tau) b(\tau) b(0) b(0) \rangle_{\rho}
    }{
    \langle b^\dagger(0) b(0) \rangle_{\rho}
    }
    =
    g_{\text{ac}}^{(3)}(0,\tau)\ \langle b^\dagger(0) b(0) \rangle_{\rho}^2
\end{equation*}
and the denominator equates to
\begin{align*}
    \frac{
    \langle b^\dagger(0) b^\dagger(0) b(0) b(0) \rangle_{\rho}
    }{
    \langle b^\dagger(0) b(0) \rangle_{\rho}
    } 
    \ 
    \frac{
    \langle b^\dagger(0) b^\dagger(\tau) b(\tau) b(0) \rangle_{\rho}
    }{
    \langle b^\dagger(0) b(0) \rangle_{\rho}
    }\\
    =
    g_{\text{ac}}^{(2)}(0)\ g_{\text{ac}}^{(2)}(\tau)\ \langle b^\dagger(0) b(0) \rangle_{\rho}^2 
\end{align*}
so that 
\begin{equation}
    g_{\text{ac}}^{(2)}\big|_{-1}(\tau) 
    = \frac{
    g_{\text{ac}}^{(3)}(0,\tau)
    }{
    g_{\text{ac}}^{(2)}(0)\ g_{\text{ac}}^{(2)}(\tau)
    }
\label{EqGMinus1}
\end{equation}
And similarly,
\begin{equation}
    h_{\text{ac}}^{(2)}\big|_{+1}(\tau) 
    = \frac{
    h_{\text{ac}}^{(3)}(0,\tau)
    }{
    h_{\text{ac}}^{(2)}(0)\ h_{\text{ac}}^{(2)}(\tau)
    }
\label{EqHPlus1}
\end{equation}

For the helium acoustic oscillator in this work that is initialized in a thermal state $\rho_\text{th}$, this gives
\begin{equation*}
    g_{\text{ac}}^{(2)}\big|_{-1,\text{th}}(\tau)
    =
    h_{\text{ac}}^{(2)}\big|_{+1,\text{th}}(\tau) 
    =
    \frac{
    1+2 e^{-\bar\gamma_\text{ac} t}
    }{
    1+e^{-\bar\gamma_\text{ac} t}
    }.
\end{equation*}
These functions are plotted as solid lines alongside the measurements in Fig.~4(b) of the main text.
Note that evaluation of the data points in Fig.~4(b) makes explicit use of post-selected data. 
This is because the experimental realization of Eq. \ref{EqGMinus1}, \ref{EqHPlus1},
\begin{align*}
    \frac{
    g_{\text{ac}}^{(3)}(0,\tau)
    }{
    g_{\text{ac}}^{(2)}(0)\ g_{\text{ac}}^{(2)}(\tau)
    }
    &=\frac{
    C^{(3)}_\text{AS}(0^+,\tau) / C^{(3)}_\text{AS}(\infty,\infty)}
    {C^{(2)}_\text{AS}(0^+) / C^{(2)}_\text{AS}(\infty)
    \cdot C^{(2)}_\text{AS}(\tau) / C^{(2)}_\text{AS}(\infty)},\\
    \frac{
    h_{\text{ac}}^{(3)}(0,\tau)
    }{
    h_{\text{ac}}^{(2)}(0)\ h_{\text{ac}}^{(2)}(\tau)}
    &=\frac{
    C^{(3)}_\text{S}(0^+,\tau) / C^{(3)}_\text{S}(\infty,\infty)}
    {C^{(2)}_\text{S}(0^+) / C^{(2)}_\text{S}(\infty)
    \cdot C^{(2)}_\text{S}(\tau) / C^{(2)}_\text{S}(\infty)}
\end{align*}
involves $C^{(3)}_\text{AS(S)}(0^+,\tau)$ (and $C^{(2)}_\text{AS(S)}(\tau)$) which are photon pairs (and counts) conditioned on the arrival of an anti-Stokes (Stokes) photon at time $t=0$, which heralds the subtraction (addition) of one phonon.
As described in Sec.~\ref{TheoryCorrelations} and Fig.~\ref{Fig10}, each $\infty$ in the normalizations $C^{(3)}_\text{AS(S)}(\infty,\infty)$ and $C^{(2)}_\text{AS(S)}(\infty)$ represents times $\tau > 10\bar\gamma_\text{ac}^{-1}$.
The time $t=0^+$ represents the finite time of $\SI{10}{\micro\second}$ (or $\SI{5}{\micro\second}$) of the experimental data bin used.
It corresponds to the time of the first click used to measure the $g^{(2)}_\text{ac}$ (or $h^{(2)}_\text{ac})$ of the phonon subtracted (or added) state.
This leads to a finite-time correction of the theoretical predictions, using Eq. \ref{EqGMinus1}, \ref{EqHPlus1}.
This finite-time correction results in a slight shift of the theoretical curves shown in Fig.~4(b); however since it is negligible on the scale of Fig.~4(b) we choose not to show it.

More generally, the  subtraction (addition) of $k$ phonons at time $t=0$ from a state described by density matrix $\rho$ yields the state with density matrix $\rho_{-k}$ ($\rho_{+k}$)
\begin{equation}
    \rho_{-k}
    =\frac{
    (b(0))^k\ \rho\ (b^\dagger(0))^k
    }{
    \langle (b^\dagger(0))^k\ (b(0))^k \rangle_{\rho}
    }
\label{EqRhoMinusK}
\end{equation}
\begin{equation}
    \rho_{+k}
    =\frac{
    (b^\dagger(0))^k\ \rho\ (b(0))^k
    }{
    \langle (b(0))^k\ (b^\dagger(0))^k \rangle_{\rho}
    }
\label{EqRhoPlusK}
\end{equation}
Through an \emph{ab-initio} evaluation similar to that illustrated above, the $n^\text{th}$-order coherence of a $k$-phonon subtracted (added) state is evaluated in terms of the coherences of the steady state $\rho$ to be
\begin{equation*}
    g_{\text{ac}}^{(n)}\big|_{-k}(
    \boldsymbol{\tau}) 
    = \frac{
    g_\text{ac}^{(k+n)} (\boldsymbol{0}^{\otimes k},\boldsymbol{\tau})\ 
    \big(g_\text{ac}^{(k)} (\boldsymbol{0})\big)^{n-1}
    }{
    g_\text{ac}^{(k+1)} (\boldsymbol{0})
    \big[\prod_{p=1}^{n-1} g_\text{ac}^{(k+1)} \big(\boldsymbol{0}^{\otimes (k-1)},t_p\big)\big]
    }
\label{generalG}
\end{equation*}
\begin{equation*}
    h_{\text{ac}}^{(n)}\big|_{+k}(
    \boldsymbol{\tau}) 
    = \frac{
    h_\text{ac}^{(k+n)} (\boldsymbol{0}^{\otimes k},\boldsymbol{\tau})\ 
    \big(h_\text{ac}^{(k)} (\boldsymbol{0})\big)^{n-1}
    }{
    h_\text{ac}^{(k+1)} (\boldsymbol{0})
    \big[\prod_{p=1}^{n-1} h_\text{ac}^{(k+1)} \big(\boldsymbol{0}^{\otimes (k-1)},t_p\big)\big]
    }
\label{generalH}
\end{equation*}
where $\boldsymbol{\tau}=(\tau_1,\tau_2,...,\tau_{n-1})$, $\boldsymbol{0}^{\otimes k}=(0,0,...\ k\ \text{times})$, and $t_p=\sum_{j=1}^p\tau_j$ is the $(p+1)^\text{th}$ time. (Recall that $\tau_k$ is the delay between the $k^\text{th}$ and $(k+1)^\text{th}$ time.)

For the helium acoustic oscillator in this work that is initialized in a thermal state $\rho_\text{th}$, then,
\begin{align*}
g^{(2)}_{\text{ac}}\vert _{-k}(\tau) &=
\frac{g^{(k+2)}_\text{ac}(\textbf{0}^{\otimes k},\tau) g^{(k)}_\text{ac}(\textbf{0})}
{g^{(k+1)}_\text{ac}(\textbf{0}) g^{(k+1)}_\text{ac}(\boldsymbol{0}^{\otimes (k-1)},\tau)},\\
g^{(2)}_{\text{ac}}\vert _{-k,\text{th}}(\tau) &=
\frac{1+(k+1)e^{-\bar\gamma_\text{ac}\tau}}{1+k e^{-\bar\gamma_\text{ac}\tau}},
\end{align*}
and
\begin{align*}
h^{(2)}_{\text{ac}}\vert _{+k}(\tau) &=
\frac{h^{(k+2)}_\text{ac}(\textbf{0}^{\otimes k},\tau) h^{(k)}_\text{ac}(\textbf{0})}
{h^{(k+1)}_\text{ac}(\textbf{0}) h^{(k+1)}_\text{ac}(\boldsymbol{0}^{\otimes (k-1)},\tau)},\\
h^{(2)}_{\text{ac}}\vert _{+k,\text{th}}(\tau) &=
\frac{1+(k+1)e^{-\bar\gamma_\text{ac}\tau}}{1+k e^{-\bar\gamma_\text{ac}\tau}}.
\end{align*}

\textbf{Mean phonon numbers} The mean phonon number of a $k$-phonon subtracted state at time $\tau$ after its generation is given by (using Eq.~\ref{EqRhoMinusK})
\begin{equation*}
    n_\text{ac}^{-k}(\tau) 
    =
    \langle b^\dagger(\tau) b(\tau) \rangle_{\rho_{-k}} 
    =
    \frac{
    \langle (b^\dagger(0))^k b^\dagger(\tau) b(\tau) (b(0))^k \rangle_{\rho}
    }{
    \langle (b^\dagger(0))^k\ (b(0))^k \rangle_{\rho}
    }.
\end{equation*}
Dividing both sides by $n_\text{ac}(0)=\langle b^\dagger(0) b(0) \rangle_{\rho}$ yields
\begin{equation}
    \frac{n_\text{ac}^{-k}(\tau)}{n_\text{ac}(0)}
    =
    \frac{
    g_\text{ac}^{(k+1)} (\boldsymbol{0}^{\otimes (k-1)},\tau)
    }{
    g_\text{ac}^{(k)} (\boldsymbol{0})
    }
\label{EqNRatioMinus}
\end{equation}
Eq.~\ref{EqNRatioMinus} has an intuitive interpretation in terms of the scattering rates of photons by the mechanical oscillator.
The numerator $g_\text{ac}^{(k+1)} (\boldsymbol{0}^{\otimes (k-1)},\tau)$ is proportional to the rate of detecting an anti-Stokes photon at time $t=\tau$ after a $k-$phonon subtraction has been heralded at time $t=0$ (by the detection of $k$-photons at time $t=0$).
The denominator $g_\text{ac}^{(k)} (\boldsymbol{0}) = g_\text{ac}^{(k+1)} (\boldsymbol{0}^{\otimes (k-1)},\infty)$ is proportional to the rate of detecting an anti-Stokes photon at time $t\rightarrow\infty$ after a $k-$phonon subtraction has been heralded at time $t=0$.
The proportionality constant for both is simply the rate of detecting uncorrelated arrivals of $k+1$ photons.
The numerator is also proportional to $n_\text{ac}^{-k}(\tau)$ (as the anti-Stokes scattering rate is proportional to the mean phonon occupancy -- the proportionality constant is simply the scattering rate $\gamma_\text{AS}$, of the main text, times the net detection efficiency).
Whereas the denominator is proportional to $n_\text{ac}^{-k}(\infty) = n_\text{ac}(0)$; the latter equality coming from the fact that for $\tau\rightarrow\infty$, the heralded $k$-phonon subtracted state equilibrates back to the steady state.
Taking a ratio of these rates, interpreted these two ways, readily yields Eq.~\ref{EqNRatioMinus}.

Similarly, the mean phonon number of a $k$-phonon added state at time $\tau$ after its generation is given by (using Eq.~\ref{EqRhoPlusK})
\begin{align*}
    n_\text{ac}^{+k}(\tau) +1
    &=
    \langle b^\dagger(\tau) b(\tau) \rangle_{\rho_{+k}} +1\\
    &=
    \langle b(\tau) b^\dagger(\tau) \rangle_{\rho_{+k}}\\
    &=
    \frac{
    \langle  (b(0))^k b(\tau) b^\dagger(\tau) (b^\dagger(0))^k \rangle_{\rho}
    }{
    \langle (b(0))^k\ (b^\dagger(0))^k \rangle_{\rho}
    }.
\end{align*}
Dividing both sides by $n_\text{ac}(0)+1=\langle b(0) b^\dagger(0) \rangle_{\rho}$ yields
\begin{equation*}
    \frac{n_\text{ac}^{+k}(\tau)+1}{n_\text{ac}(0)+1}
    =
    \frac{
    h_\text{ac}^{(k+1)}(\boldsymbol{0}^{\otimes (k-1)},\tau)
    }{
    h_\text{ac}^{(k)} (\boldsymbol{0})
    }.
\end{equation*}
This equation has a similar intuitive interpretation as Eq.~\ref{EqNRatioMinus}, but with the rates now being proportional to the Stokes scattering rates, which are proportional to $n_\text{ac}+1$.

Note again that the experimental evaluation of the data points in Fig.~4(a) involves the explicit use of post-selected data. 
This is because the evaluation
\begin{align*}
    \frac{
    g_\text{ac}^{(k+1)} (\boldsymbol{0}^{\otimes (k-1)},\tau)
    }{
    g_\text{ac}^{(k+1)} (\boldsymbol{0}^{\otimes (k-1)},\infty)
    }
    &=\frac{
    C^{(k+1)}_\text{AS}(\boldsymbol{0}^{\otimes (k-1)},\tau) / C^{(k+1)}_\text{AS}(\boldsymbol{\infty}^{\otimes k})}
    { C^{(k+1)}_\text{AS}(\boldsymbol{0}^{\otimes (k-1)},\infty) / C^{(k+1)}_\text{AS}(\boldsymbol{\infty}^{\otimes k})},\\
    \frac{
    h_\text{ac}^{(k+1)} (\boldsymbol{0}^{\otimes (k-1)},\tau)
    }{
    h_\text{ac}^{(k+1)} (\boldsymbol{0}^{\otimes (k-1)},\infty)
    }
    &=\frac{
    C^{(k+1)}_\text{S}(\boldsymbol{0}^{\otimes (k-1)},\tau) / C^{(k+1)}_\text{S}(\boldsymbol{\infty}^{\otimes k})}
    { C^{(k+1)}_\text{S}(\boldsymbol{0}^{\otimes (k-1)},\infty) / C^{(k+1)}_\text{S}(\boldsymbol{\infty}^{\otimes k})}
\end{align*}
involves $C^{(k+1)}_\text{AS(S)}(\boldsymbol{0}^{\otimes (k-1)},\tau)$ which are photon counts conditioned on the arrival of $k$ anti-Stokes (Stokes) photons at time $t=0$, which herald the subtraction (addition) of $k$ phonons.
As described earlier, each $\infty$ in the normalizations represents times $\tau > 10\bar\gamma_\text{ac}^{-1}$.

\bibliography{supp}